\documentclass[11pt, a4paper]{article}

\makeatletter%

\usepackage{ifthen}



\makeatletter%

\makeatletter
\newcommand{\ifndef}[2]{\@ifundefined{#1}{#2}{}}
\makeatother

\newcommand{\mydef}[2]{\def#1{#2}}

\newcommand{\nospell}[1]{#1}  %

\makeatletter
\newcommand{\myusepackage}[2][]{\@ifpackageloaded{#2}{} %
{\ifthenelse{\equal{}{#1}} {\usepackage{#2}} {\usepackage[#1]{#2}} }}
\makeatother

\myusepackage{amssymb}
\myusepackage{amsmath}  %

\myusepackage{amsthm}  %

{}
\myusepackage[T1,OT2,T2A]{fontenc}
\DeclareTextSymbolDefault{\CYRYAT}{OT2}
\DeclareTextSymbolDefault{\cyryat}{OT2}
\DeclareTextSymbolDefault{\CYRFITA}{OT2}
\DeclareTextSymbolDefault{\cyrfita}{OT2}
\DeclareTextSymbolDefault{\CYRIZH}{OT2}
\DeclareTextSymbolDefault{\cyrizh}{OT2}
\myusepackage[utf8]{inputenc}
\let\f\relax

{}
\usepackage[greek,ukrainian,russian,english]{babel}
{}

\myusepackage{latexsym}
\myusepackage{amsfonts}
\myusepackage{units}    %
\myusepackage{psfrag}   %
\myusepackage{url}   %
\myusepackage{hyperref}  %
\setpdflinkmargin{0.75pt}

\usepackage[usenames,dvipsnames]{xcolor}
\hypersetup{
colorlinks,
linkcolor={red!55!black},
citecolor={blue!55!black},
urlcolor={blue!35!black}
}

\myusepackage{hyphenat}  %
\myusepackage{caption}   %
\myusepackage{microtype}   %
\myusepackage{marginnote}  %
\myusepackage{calc}  %
{}  %
\myusepackage{enumitem}  %
{}
\myusepackage[dvips]{graphicx}
{}
{}  %
{} %

{}     %
\newcommand{\dgCapDefinition}{Definition}
\newcommand{\dgCapDefinitions}{Definitions}
\newcommand{\dgCapPostulate}{Postulate}
\newcommand{\dgCapPostulates}{Postulates}
\newcommand{\dgCapExample}{Example}

\newcommand{\dgCapFact}{Fact}
\newcommand{\dgCapFacts}{Facts}
\newcommand{\dgCapQuestion}{Question}
\newcommand{\dgCapQuestions}{Questions}
\newcommand{\dgCapLemma}{Lemma}
\newcommand{\dgCapLemmas}{Lemmas}

\newcommand{\dgCapCorollary}{Corollary}
\newcommand{\dgCapCorollaries}{Corollaries}
\newcommand{\dgCapProposition}{Proposition}
\newcommand{\dgCapPropositions}{Propositions}
\newcommand{\dgCapClaim}{Claim}
\newcommand{\dgCapClaims}{Claims}
\newcommand{\dgCapTheorem}{Theorem}
\newcommand{\dgCapTheorems}{Theorems}
\newcommand{\dgCapProblem}{Problem}
\newcommand{\dgCapProblems}{Problems}
\newcommand{\dgCapRemark}{Remark}
\newcommand{\dgCapRemarks}{Remarks}
\newcommand{\dgCapConjecture}{Conjecture}
\newcommand{\dgCapConjectures}{Conjectures}
\newcommand{\dgCapResult}{Result}

\newcommand{\dgCapChapter}{Chapter}
\newcommand{\dgCapChapters}{Chapters}
\newcommand{\dgCapSection}{Section}
\newcommand{\dgCapSections}{Sections}
\newcommand{\dgCapSubsection}{Subsection}
\newcommand{\dgCapSubsections}{Subsections}
\newcommand{\dgCapFigure}{Figure}
\newcommand{\dgCapFigures}{Figures}
\newcommand{\dgCapEquation}{Equation}
\newcommand{\dgCapEquations}{Equations}
\newcommand{\dgCapExpression}{Expression}
\newcommand{\dgCapExpressions}{Expressions}
\newcommand{\dgCapInequality}{Inequality}
\newcommand{\dgCapInequalities}{Inequalities}
\newcommand{\dgProofOf}{\proofname\ of}
{}
\newcommand{\dgDefinition}{Definition}
\newcommand{\dgDefinitions}{Definitions}
\newcommand{\dgPostulate}{Postulate}
\newcommand{\dgPostulates}{Postulates}

\newcommand{\dgFact}{Fact}
\newcommand{\dgFacts}{Facts}
\newcommand{\dgQuestion}{Question}
\newcommand{\dgQuestions}{Questions}
\newcommand{\dgLemma}{Lemma}
\newcommand{\dgLemmas}{Lemmas}
\newcommand{\dgCorollary}{Corollary}
\newcommand{\dgCorollaries}{Corollaries}
\newcommand{\dgProposition}{Proposition}
\newcommand{\dgPropositions}{Propositions}
\newcommand{\dgClaim}{Claim}
\newcommand{\dgClaims}{Claims}
\newcommand{\dgTheorem}{Theorem}
\newcommand{\dgTheorems}{Theorems}
\newcommand{\dgProblem}{Problem}
\newcommand{\dgProblems}{Problems}
\newcommand{\dgRemark}{Remark}
\newcommand{\dgRemarks}{Remarks}
\newcommand{\dgConjecture}{Conjecture}
\newcommand{\dgConjectures}{Conjectures}

\newcommand{\dgChapter}{Chapter}
\newcommand{\dgChapters}{Chapters}
\newcommand{\dgSection}{Section}
\newcommand{\dgSections}{Sections}
\newcommand{\dgSubsection}{Subsection}
\newcommand{\dgSubsections}{Subsections}
\newcommand{\dgFigure}{Figure}
\newcommand{\dgFigures}{Figures}
\newcommand{\dgEquation}{Equation}
\newcommand{\dgEquations}{Equations}
\newcommand{\dgExpression}{Expression}
\newcommand{\dgExpressions}{Expressions}
\newcommand{\dgInequality}{Inequality}
\newcommand{\dgInequalities}{Inequalities}
{}
{}

{}
\ifndef{theorem}{}
\ifndef{theorem*}{\newtheorem*{theorem*}{\dgCapTheorem}}
{}

{}  %
{}  %
\ifndef{lemma}{}
\ifndef{lemma*}{\newtheorem*{lemma*}{\dgCapLemma}}
\ifndef{corollary}{}
\ifndef{corollary*}{\newtheorem*{corollary*}{\dgCapCorollary}}
\ifndef{conjecture}{}
\ifndef{conjecture*}{\newtheorem*{conjecture*}{\dgCapConjecture}}
\ifndef{proposition}{}
\ifndef{proposition*}{\newtheorem*{proposition*}{\dgCapProposition}}
\ifndef{claim}{}
\ifndef{claim*}{\newtheorem*{claim*}{\dgCapClaim}}
\ifndef{result}{}
\ifndef{result*}{\newtheorem*{result*}{\dgCapResult}}
\ifndef{problem}{}
\ifndef{problem*}{\newtheorem*{problem*}{\dgCapProblem}}
{}  %
{}  %

\newtheoremstyle{mydefinition}  %
{\topsep}{\topsep}  %
{\slshape}  %
{}  %
{\bfseries}  %
{.}  %
{ }  %
{}  %

\newtheoremstyle{myremark}  %
{\topsep}{\topsep}  %
{\slshape}  %
{}  %
{\bfseries\itshape}  %
{.}  %
{ }  %
{\thmname{#1}\thmnumber{ \!#2}}  %

\newtheoremstyle{myexample}  %
{\topsep}{\topsep}  %
{\itshape}  %
{}  %
{\slshape}  %
{.}  %
{ }  %
{\ul{\thmname{#1}}}  %

\newtheoremstyle{myclaims}  %
{\topsep}{\topsep}  %
{\slshape}  %
{}  %
{\bfseries\slshape}  %
{.}  %
{ }  %
{\thmname{#1}\thmnumber{ \!#2}\ifthenelse{\equal{}{#3}}%
{}{\textnormal{ \!(#3)}}}  %

{} %
{} %
{\theoremstyle{myremark}

\newtheorem*{myremark*}{\dgCapRemark}
}
{} %
{\theoremstyle{mydefinition}
\ifndef{postulate}{}
\ifndef{postulate*}{\newtheorem*{postulate*}{\dgCapPostulate}}
\ifndef{definition}{}
\ifndef{definition*}{\newtheorem*{definition*}{\dgCapDefinition}}
}
{\theoremstyle{myexample}
\ifndef{example}{}
\ifndef{example*}{\newtheorem*{example*}{\dgCapExample}}
}
{\theoremstyle{myclaims}

\newtheorem*{my_claim*}{\dgCapClaim}
\ifndef{fact}{}
\ifndef{fact*}{\newtheorem*{fact*}{\dgCapFact}}
\ifndef{question}{}
\ifndef{question*}{\newtheorem*{question*}{\dgCapQuestion}}
}
{} %
{} %
{} %

{}
\newtheoremstyle{anystatementst}  %
{\topsep}{\topsep}  %
{\itshape}  %
{}  %
{\bfseries}  %
{.}  %
{ }  %
{#3}  %

{\theoremstyle{anystatementst} }

{}

\newcommand{\MyUniPat}{lsdfgkhjvrkjlhmisdlcjn}

\newcommand{\newident}[3][\MyUniPat]{\ifthenelse{\equal{\MyUniPat}{#1}}
{
\newcommand{#2}[1][]{\Ensuremath{\mathit{#3##1}}}
}
{\ifthenelse{\equal{}{#1}}
{
\newcommand{#2}[1][]{\Ensuremath{\mathit{#3}}}
}
{
\newcommand{#2}[1][\MyUniPat]{\ifthenelse{\equal{\MyUniPat}{##1}}%
{\Ensuremath{\mathit{#1}}}%
{\Ensuremath{\mathit{#3}}}}
}
}
}

\newcommand{\newidenT}[3][\MyUniPat]{\ifthenelse{\equal{\MyUniPat}{#1}}
{
\newcommand{#2}[1][\MyUniPat]{\ifthenelse{\equal{\MyUniPat}{##1}}%
{\il{#3}}%
{\Ensuremath{\mathit{#3##1}}}}
}
{
\newcommand{#2}[1][\MyUniPat]{\ifthenelse{\equal{\MyUniPat}{##1}}%
{\il{#1}}%
{\Ensuremath{\mathit{#3}}}}
}
}

\newcommand{\newmat}[3][\MyUniPat]{\ifthenelse{\equal{\MyUniPat}{#1}}%
{\newcommand{#2}[1][]{\Ensuremath{#3##1}}}%
{\newcommand{#2}[1][]{\Ensuremath{#3}}}%
}

\newcommand{\providemat}[3][\MyUniPat]{\ifthenelse{\equal{\MyUniPat}{#1}}
{\providecommand{#2}[1][]{\Ensuremath{#3##1}}}
{\providecommand{#2}[1][]{\Ensuremath{#3}}}  %
}

\newcommand{\newmatop}[3][\MyUniPat]{\ifthenelse{\equal{\MyUniPat}{#1}}
{
\mydef{#2}{\operatorname{#3}}
}
{
\newcommand{#2}[1][\MyUniPat]{\ifthenelse{\equal{\MyUniPat}{##1}}%
{\operatorname{#1}}%
{\operatorname{#3}}}
}
}

\newcommand{\newfunction}[2]{%
\newcommand{#1}[2][\MyUniPat]{\ifthenelse{\equal{\MyUniPat}{##1}}%
{\Ensuremath{#2\lf(##2\rt)}}%
{#2(##2)}}%
}

\makeatletter
\newcommand{\MyMakeTheoMacros}[3]{
\expandafter\newcommand\csname\expandafter\@gobble\string#2NostarNoname@DGaux\endcsname[2][]
{\ifthenelse{\equal{}{##1}}%
{\begin{#1}~##2 \end{#1}}%
{\begin{#1}\label{##1}~##2\end{#1}}%
}
\expandafter\newcommand\csname\expandafter\@gobble\string#2StarNoname@DGaux\endcsname[1]
{\begin{#1*}~##1 \end{#1*}}
\def#2{\expandafter\@ifstar%
\expandafter{\csname\expandafter\@gobble\string#2StarNoname@DGaux\endcsname}%
{\csname\expandafter\@gobble\string#2NostarNoname@DGaux\endcsname}%
}

\expandafter\newcommand\csname\expandafter\@gobble\string#2NostarName@DGaux\endcsname[3][]
{\ifthenelse{\equal{}{##1}}%
{\begin{#1}[\e{##2}]~##3 \end{#1}}%
{\begin{#1}[\e{##2}]\label{##1}~##3\end{#1}}%
}
\expandafter\newcommand\csname\expandafter\@gobble\string#2StarName@DGaux\endcsname[2]
{\begin{#1*}[\e{##1}]~##2 \end{#1*}}
\def#3{\expandafter\@ifstar%
\expandafter{\csname\expandafter\@gobble\string#2StarName@DGaux\endcsname}
{\csname\expandafter\@gobble\string#2NostarName@DGaux\endcsname}%
}
}
\makeatother

\makeatletter
\newtheorem*{rep@theorem}{\rep@title}
\newcommand{\newreptheorem}[2]{%
\newenvironment{rep#1}[1]{%
\def\rep@title{#2 \ref{##1}}%
\begin{rep@theorem}}%
{\end{rep@theorem}}}
\makeatother

\newcommand{\MyMakeDupTheoMacros}[7]{
\MyMakeTheoMacros{#1}{#2}{#3}
\newreptheorem{#1}{#6}
\newcommand{#4}[3]{
\newcommand{##2}{##3}
\begin{#1}\label{##1}~##2\end{#1}}
\newcommand{#5}[4]{
\newcommand{##2}{##4}
\begin{#1}{\e{##3}}\label{##1}~##2\end{#1}}
\newcommand{#7}[2]{\begin{rep#1}{##1}~##2 \end{rep#1}}
}

{}
\newcommand{\MyMakeRefMacros}[3]{\newcommand{#1}[2][]
{\ifthenelse{\equal{}{##1}}{#2~\ref{##2}}{#3~\ref{##1} and~\ref{##2}}}}

\newcommand{\MyMakeEqRefMacros}[3]{\newcommand{#1}[2][]
{\ifthenelse{\equal{}{##1}}{#2~\eqref{##2}}{#3~\eqref{##1} and~\eqref{##2}}}}
{}

{}  %
\newcommand{\bibentry}[8]{
{}\bibitem[\nospell{#8}]{#1} {\textup #3}.{}
\ifthenelse{\equal{}{#6}}
{\newblock \textrm{#4.} \newblock {\em #5}, #7....}
{\newblock \textrm{#4.} \newblock {\em #5, #6}, #7.}
}

{} %

\MyMakeTheoMacros{fact}{\fct}{\nfct}

\MyMakeRefMacros{\fctref}{\dgFact}{\dgFacts}
\MyMakeRefMacros{\Fctref}{\dgCapFact}{\dgCapFacts}

\MyMakeTheoMacros{question}{\quest}{\nquest}

\MyMakeRefMacros{\questref}{\dgQuestion}{\dgQuestions}
\MyMakeRefMacros{\Questref}{\dgCapQuestion}{\dgCapQuestions}

{} %
\MyMakeDupTheoMacros{lemma}
{\lem}{\nlem}{\lemdup}{\nlemdup}{\dgLemma}{\lemrep}
{}

\MyMakeRefMacros{\lemref}{\dgLemma}{\dgLemmas}
\MyMakeRefMacros{\Lemref}{\dgCapLemma}{\dgCapLemmas}

\newcommand{\fakelemref}[1]{\dgLemma~{#1}}

\MyMakeDupTheoMacros{corollary}
{\crl}{\ncrl}{\crldup}{\ncrldup}{\dgCorollary}{\crlrep}

\MyMakeRefMacros{\crlref}{\dgCorollary}{\dgCorollaries}
\MyMakeRefMacros{\Crlref}{\dgCapCorollary}{\dgCapCorollaries}

\MyMakeTheoMacros{proposition}{\prp}{\nprp}

{}
\newtheorem*{prp*}{\e{\dgCapProposition}}

{}

\MyMakeRefMacros{\prpref}{\dgProposition}{\dgPropositions}
\MyMakeRefMacros{\Prpref}{\dgCapProposition}{\dgCapPropositions}

\MyMakeDupTheoMacros{my_claim}
{\clm}{\nclm}{\clmdup}{\nclmdup}{\dgClaim}{\clmrep}

\MyMakeRefMacros{\clmref}{\dgClaim}{\dgClaims}
\MyMakeRefMacros{\Clmref}{\dgCapClaim}{\dgCapClaims}

\MyMakeDupTheoMacros{theorem}
{\theo}{\ntheo}{\theodup}{\ntheodup}{\dgTheorem}{\theorep}

\MyMakeRefMacros{\theoref}{\dgTheorem}{\dgTheorems}
\MyMakeRefMacros{\Theoref}{\dgCapTheorem}{\dgCapTheorems}

\MyMakeTheoMacros{postulate}{\postu}{\npostu}

\MyMakeRefMacros{\posturef}{\dgPostulate}{\dgPostulates}
\MyMakeRefMacros{\Posturef}{\dgCapPostulate}{\dgCapPostulates}

\MyMakeTheoMacros{definition}{\defi}{\ndefi}

\MyMakeRefMacros{\defiref}{\dgDefinition}{\dgDefinitions}
\MyMakeRefMacros{\Defiref}{\dgCapDefinition}{\dgCapDefinitions}

\MyMakeTheoMacros{problem}{\prob}{\nprob}

\MyMakeRefMacros{\probref}{\dgProblem}{\dgProblems}
\MyMakeRefMacros{\Probref}{\dgCapProblem}{\dgCapProblems}

\MyMakeTheoMacros{myremark}{\rem}{\nrem}

\MyMakeRefMacros{\remref}{\dgRemark}{\dgRemarks}
\MyMakeRefMacros{\Remref}{\dgCapRemark}{\dgCapRemarks}

\newcommand{\fakeremref}[1]{\dgRemark~{#1}}

\MyMakeTheoMacros{conjecture}{\conj}{\nconj}

\MyMakeRefMacros{\conjref}{\dgConjecture}{\dgConjectures}
\MyMakeRefMacros{\Conjref}{\dgCapConjecture}{\dgCapConjectures}

\renewcommand{\qedsymbol}{$\blacksquare$}

\newcommand{\prf}[2][]{\ifthenelse{\equal{}{#1}}%
{\begin{proof}\renewcommand{\qedsymbol}{$\blacksquare$}%
#2 \end{proof}}%
{\begin{proof}[\dgProofOf\ #1]%
\renewcommand{\qedsymbol}{$\blacksquare_{\mbox{\it{\scriptsize{#1}}}}$}%
#2 \end{proof}\renewcommand{\qedsymbol}{$\blacksquare$}}%
}

\newcommand{\prfstart}[1][]{\ifthenelse{\equal{}{#1}}%
{\begin{proof}\renewcommand{\qedsymbol}{$\blacksquare$}}%
{\begin{proof}[\dgProofOf\ #1]%
\renewcommand{\qedsymbol}{$\blacksquare_{\mbox{\it{\scriptsize{#1}}}}$}}%
}
\newcommand{\prfend}[1][*]{%
\ifthenelse{\equal{}{#1}}{\renewcommand{\qedsymbol}{$\blacksquare$}}{}%
\ifthenelse{\equal{*}{#1}}{}%
{\renewcommand{\qedsymbol}{$\blacksquare_{\mbox{\it{\scriptsize{#1}}}}$}}%
\end{proof}\renewcommand{\qedsymbol}{$\blacksquare$}%
}

\makeatletter

\newcommand{\NewcommandThreeArgsTwoOpt}[5]{
\DeclareRobustCommand#1{\@ifnextchar[%
{\csname\expandafter\@gobble\string#1@presq\endcsname}%
{\csname\expandafter\@gobble\string#1@nopresq\endcsname}}
\expandafter\def\csname\expandafter\@gobble\string#1@nopresq\endcsname##1{\@ifnextchar[%
{\csname\expandafter\@gobble\string#1@nopresq@postsq\endcsname[]{##1}}%
{\csname\expandafter\@gobble\string#1@nopresq@nopostsq\endcsname[]{##1}}}
\expandafter\def\csname\expandafter\@gobble\string#1@presq\endcsname[##1]##2{\@ifnextchar[%
{\csname\expandafter\@gobble\string#1@presq@postsq\endcsname[{##1}]{##2}}%
{\csname\expandafter\@gobble\string#1@presq@nopostsq\endcsname[{##1}]{##2}}}
\expandafter\def\csname\expandafter\@gobble\string#1@nopresq@nopostsq\endcsname[##1]##2{#2}
\expandafter\def\csname\expandafter\@gobble\string#1@presq@nopostsq\endcsname[##1]##2{#3}
\expandafter\def\csname\expandafter\@gobble\string#1@nopresq@postsq\endcsname[##1]##2[##3]{#4}
\expandafter\def\csname\expandafter\@gobble\string#1@presq@postsq\endcsname[##1]##2[##3]{#5}
}

\makeatother

\newcommand{\NewSectLikeDG}[2]{
\NewcommandThreeArgsTwoOpt{#1}
{\ifthenelse{\equal{*}{##2}}{#2*}{#2{##2}}}
{#2{##2}\label{##1}}
{#2{\texorpdfstring{##2}{##3}}}
{#2{\texorpdfstring{##2}{##3}}\label{##1}}
}

\NewSectLikeDG{\sect}{\section}

\NewSectLikeDG{\ssect}{\subsection}

\NewSectLikeDG{\sssect}{\subsubsection}

\MyMakeRefMacros{\chref}{\dgChapter}{\dgChapters}
\MyMakeRefMacros{\Chref}{\dgCapChapter}{\dgCapChapters}

\MyMakeRefMacros{\sref}{\dgSection}{\dgSections}
\MyMakeRefMacros{\Sref}{\dgCapSection}{\dgCapSections}

\MyMakeRefMacros{\ssref}{\dgSubsection}{\dgSubsections}
\MyMakeRefMacros{\Ssref}{\dgCapSubsection}{\dgCapSubsections}

\MyMakeRefMacros{\sssref}{\dgSubsection}{\dgSubsections}
\MyMakeRefMacros{\Sssref}{\dgCapSubsection}{\dgCapSubsections}

\MyMakeRefMacros{\figref}{\dgFigure}{\dgFigures}
\MyMakeRefMacros{\Figref}{\dgCapFigure}{\dgCapFigures}

\newcommand{\IfMathMode}[2]{\ifmmode{#1}\else{#2}\fi}

\newcommand{\Ensuremath}{\ensuremath}

\newcommand{\fbr}[1]{\IfMathMode%
{#1}{$#1$}}                     %

\newcommand{\fnbr}[1]{\mbox{\fbr{#1}}}  %

\newcommand{\fla}[2][*]{\ifthenelse{\equal{}{#1}}{\fbr{#2}}{\fnbr{#2}}}
\newcommand{\f}{\fla}

\newcommand{\bfla}[2][]{\mbox{\fbr{\pmb{#2}}}}
\newcommand{\fb}{\bfla}

\newcommand{\malabel}[1]{\addtocounter{equation}{1}\tag{\theequation}\label{#1}}

\newcommand{\mal}[2][]{\MyChangeMathMargins\delimiterfactor=1001%
\ifthenelse{\equal{}{#1}}%
{\begin{align*} #2 \end{align*}}%
{\ifthenelse{\equal{P}{#1}}%
{\allowdisplaybreaks\begin{align*} #2%
\end{align*}\interdisplaylinepenalty=10000}%
{\begin{align}\begin{split}\label{#1} #2 \end{split}\end{align}}%
}\delimiterfactor=901%
}

\newcommand{\m}{\mal}

\newcommand{\mac}{\substack}

\MyMakeEqRefMacros{\equref}{\dgEquation}{\dgEquations}
\MyMakeEqRefMacros{\Equref}{\dgCapEquation}{\dgCapEquations}

\MyMakeEqRefMacros{\expref}{\dgExpression}{\dgExpressions}
\MyMakeEqRefMacros{\Expref}{\dgCapExpression}{\dgCapExpressions}

\MyMakeEqRefMacros{\inequref}{\dgInequality}{\dgInequalities}
\MyMakeEqRefMacros{\Inequref}{\dgCapInequality}{\dgCapInequalities}

\newcommand{\bref}[1]{(\ref{#1})}

\newcommand{\twocase}[4]%
{\begin{cases} #1 &\txt{#2}\\ #3 &\txt{#4}\end{cases}}

\newcommand{\thrcase}[6]%
{\begin{cases} #1 &\txt{#2}\\ #3 &\txt{#4}\\ #5 &\txt{#6}\end{cases}}

\newcommand{\lf}{\mathopen{}\mathclose\bgroup\left}
\newcommand{\rt}{\aftergroup\egroup\right}

\providecommand{\middle}{\big}
\newcommand{\md}{\middle}

\newcommand{\chs}{\genfrac(){0cm}{}}  %

\newmatop[disc]{\disc}{disc_{#1}}

\makeatletter
\def\moverlay{\mathpalette\mov@rlay}
\def\mov@rlay#1#2{\leavevmode\vtop{%
\baselineskip\z@skip \lineskiplimit-\maxdimen
\ialign{\hfil$\m@th#1##$\hfil\cr#2\crcr}}}
\makeatother

\newcommand{\h}[2][]{\ifthenelse{\equal{}{#2}}%
{\mathop H_{#1}}%
{\mathop H_{#1}{\l({#2}\r)}}}
\newcommand{\hh}[3][]{\mathop H_{#1}%
{\l({#2}\vphantom{|_1^1}\md|\vphantom{|_1^1}{#3}\r)}}
\newcommand{\hm}[2][]{\ifthenelse{\equal{}{#2}}%
{\mathop {H_{\txt{min}}}_{#1}}%
{\mathop {H_{\txt{min}}}_{#1}{\l({#2}\r)}}}
\newcommand{\hmm}[3][]{\mathop {H_{\txt{min}}}_{#1}%
{\l({#2}\vphantom{|_1^1}\md|\vphantom{|_1^1}{#3}\r)}}

\newcommand{\KL}[2]{d_{KL}\lf({#1}\md\|\vphantom{|_1^1}{#2}\rt)}

\newcommand{\hbin}[2][]{h_2#1\l(#2\r)}        %

\providecommand{\E}[2][]{\mathop{\pmb{E}}_{#1}\lf[{#2}\rt]}

\newcommand{\PR}[2][]{\mathop{\pmb{Pr}}_{#1}\lf[{#2}\rt]}
\newcommand{\PRr}[3][]{\mathop{\pmb{Pr}}_{#1}\lf[{#2}\vphantom{|_1^1}\md|\vphantom{|_1^1}{#3}\rt]}

\renewcommand{\U}[1][]{\ifthenelse{\equal{}{#1}}%
{{\cal U}}%
{{\cal U}_{#1}}}

\providemat{\QQ}{\mathbb{Q}}
\providemat{\NN}{\mathbb{N}}
\providemat{\CC}{\mathbb{C}}
\providemat{\RR}{\mathbb{R}}
\providemat{\ZZ}{\mathbb{Z}}

\newcommand{\wbr}{\overline}   %

\newcommand{\pss}[1][]{\nospell{\ifthenelse{\equal{}{#1}}%
{\txt{'s}}%
{\fla{#1\txt{'s}}}}}

\newcommand{\pl}[1][]{\nospell{\ifthenelse{\equal{}{#1}}%
{\mskip-6mu\stackrel{\text-}{}\mskip-4mu\txt{s}}%
{\fla{#1\mskip-6mu\stackrel{\text-}{}\mskip-4mu\txt{s}}}}}

\newcommand{\ord}[1][]{\nospell{\ifthenelse{\equal{}{#1}}%
{\txt{'th}}%
{\ifthenelse{\equal{1}{#1}}{$1\txt{'st}$}{\ifthenelse{\equal{2}{#1}}{$2\txt{'nd}$}{\ifthenelse{\equal{3}{#1}}{$3\txt{'rd}$}{\fla{#1\txt{'th}}}}}}}}

\newcommand{\fr}[3][*]{%
\ifthenelse{\equal{*}{#1}}%
{\frac{#2}{#3}}{}%
\ifthenelse{\equal{/}{#1}}%
{\nicefrac{#2}{#3}}{}%
\ifthenelse{\equal{}{#1}}%
{\lf.#2\md/#3\rt.}{}%
\ifthenelse{\equal{p_}{#1}}%
{\lf.\lf(#2\rt)\md/#3\rt.}{}%
\ifthenelse{\equal{_p}{#1}}%
{\lf.#2\md/\lf(#3\rt)\rt.}{}%
\ifthenelse{\equal{pp}{#1}}%
{\lf.\lf(#2\rt)\md/\lf(#3\rt)\rt.}{}%
}

\newcommand{\dr}{\nicefrac}

\newcommand{\sq}{\sqrt}

\newcommand{\set}[2][]{\ifthenelse{\equal{}{#1}}%
{\Ensuremath{\lf\{#2\rt\}}}%
{\Ensuremath{\lf\{#2\vphantom{|_1^1}\md|\vphantom{|_1^1}#1\rt\}}}}

\newcommand{\sett}[2]{\Ensuremath{\lf\{#1\vphantom{|_1^1}\md|\vphantom{|_1^1}#2\rt\}}}

\newcommand{\Min}[2][]{\ifthenelse{\equal{}{#1}}%
{\Ensuremath{\min\lf\{#2\rt\}}}%
{\Ensuremath{\min\lf\{#2\vphantom{|_1^1}\md|\vphantom{|_1^1}#1\rt\}}}}

\newcommand{\Minn}[3][]{\ifthenelse{\equal{}{#1}}%
{\Ensuremath{\min_{#2}\lf\{#3\rt\}}}%
{\Ensuremath{\min_{#2}\lf\{#3\vphantom{|_1^1}\md|\vphantom{|_1^1}#1\rt\}}}}

\newcommand{\Maxx}[3][]{\ifthenelse{\equal{}{#1}}%
{\Ensuremath{\max_{#2}\lf\{#3\rt\}}}%
{\Ensuremath{\max_{#2}\lf\{#3\vphantom{|_1^1}\md|\vphantom{|_1^1}#1\rt\}}}}

\newfunction{\asO}{O}
\newfunction{\aso}{o}
\newfunction{\asOm}{\Omega}
\newfunction{\asT}{\Theta}

\providecommand{\ket}[1]{\Ensuremath{\lf|#1\rra}}

\newcommand{\bket}[3][]{\ifthenelse{\equal{}{#1}}%
{\Ensuremath{\lla #2\md|#3\rra}}%
{\Ensuremath{\lla #2\md|#1\md|#3\rra}}
}

\providecommand{\ip}[2]{\Ensuremath{\lla #1\,,\,#2\rra}}

\newcommand{\sz}[2][]{\ifthenelse{\equal{}{#1}}%
{\Ensuremath{\lf|#2\rt|}}%
{\Ensuremath{\lf|#2\rt|_{#1}}}}
\providecommand{\norm}[2][]{\ifthenelse{\equal{}{#1}}%
{\Ensuremath{\lf\|#2\rt\|}}%
{\Ensuremath{\lf\|#2\rt\|_{#1}}}}

\newcommand{\lra}[2][*]{\ifthenelse{\equal{}{#1}}%
{\langle #2 \rangle}%
{\lla #2 \rra}}
\providecommand{\ceil}[2][*]{\ifthenelse{\equal{}{#1}}%
{\lceil #2 \rceil}%
{\llc #2 \rrc}}
\providecommand{\floor}[2][*]{\ifthenelse{\equal{}{#1}}%
{\lfloor #2 \rfloor}%
{\llf #2 \rrf}}

\newcommand{\txt}[1]{\textrm{#1}}  %

\newcommand{\Cl}{\mathcal}  %

\DeclareMathAlphabet{\mathlowcal}{OT1}{pzc}{m}{it}

\newident[]{\R}{\mathcal R_{#1}}
\newident[]{\RIIp}{\mathcal R_{#1}^{\parallel,pub}}
\newident[]{\RIIe}{\mathcal R_{#1}^{\parallel,ent}}
\newident[]{\QII}{\mathcal Q_{#1}^{\parallel}}
\newident[]{\QIIp}{\mathcal Q_{#1}^{\parallel,pub}}
\newident[]{\QIIe}{\mathcal Q_{#1}^{\parallel,ent}}

\newidenT{\Pp}{P}
\newidenT[BPP]{\BPP}{BPP_{#1}}
\newidenT{\ZPP}{ZPP}
\newidenT{\SBP}{SBP}
\newidenT{\coSBP}{coSBP}
\newidenT[RP]{\RP}{RP_{#1}}
\newidenT{\PP}{PP}
\newidenT{\UPP}{UPP}
\newidenT{\coRP}{coRP}
\newidenT{\BQP}{BQP}
\newidenT{\NP}{NP}
\newidenT{\coNP}{coNP}
\newidenT{\AM}{AM}
\newidenT[MA]{\MA}{MA_{#1}}
\newidenT[QMA]{\QMA}{QMA_{#1}}
\newidenT[ZAM]{\ZAM}{ZAM_{#1}}
\newidenT[UAM]{\UAM}{UAM_{#1}}
\newidenT{\PH}{PH}
\newidenT{\PSPACE}{PSPACE}
\newidenT{\EXP}{EXP}
\newidenT{\NEXP}{NEXP}

\newidenT{\DNF}{DNF}
\newidenT{\Eq}{Eq}
\newidenT{\Disj}{Disj}
\newidenT{\IP}{IP}

\newcommand{\lla}{\lf\langle}
\newcommand{\rra}{\rt\rangle}
\newcommand{\llc}{\lf\lceil}
\newcommand{\rrc}{\rt\rceil}
\newcommand{\llf}{\lf\lfloor}
\newcommand{\rrf}{\rt\rfloor}

\newcommand{\Then}{\Longrightarrow}

\newcommand{\tm}{\cdot}
\newcommand{\xor}{\oplus}
\newcommand{\sbseq}{\subseteq}
\newcommand{\smin}{\setminus}
\newcommand{\eps}{\varepsilon}
\newcommand{\deq}{\stackrel{\textrm{def}}{=}}

\newcommand{\unin}{\mathrel{\subset\mkern-13.1mu\sim}}  %

\newcommand{\ds}[1][]
{\ifthenelse{\equal{}{#1}}{\allowbreak\dots}{#1\allowbreak\dots#1}}
\newmat{\dc}{\ds[,]}
\newmat{\dpl}{\ds[+]}
\newmat{\dtm}{\cdots}

\mydef{\01}{\set{0,1}}
\mydef{\12}{\set{1,2}}

\mydef{\l(}{\lf(}
\mydef{\r)}{\rt)}

\mydef{\=}{\equiv}

\mathchardef\myhyphen="2D
\mydef{\-}{\myhyphen}

\newcommand{\abstr}[1]{\begin{abstract} #1 \end{abstract}}

{} %
\newcommand{\itemi}[2][\MyUniPat]{\ifthenelse{\equal{\MyUniPat}{#1}}%
{\begin{itemize}[noitemsep,topsep=3pt] #2 \end{itemize}}%
{\begin{itemize}[#1] #2 \end{itemize}}}

\newcommand{\itstart}[1][\MyUniPat]{\ifthenelse{\equal{\MyUniPat}{#1}}%
{\begin{itemize}[noitemsep,topsep=3pt]}%
{\begin{itemize}[#1]}%
}

{}  %

\newcommand{\itend}{\end{itemize}}

\protected \def \dg #1{%
\textcolor{Red}
{
{\normalmarginpar\marginnote{\bl{DG's comment}}}
{\reversemarginpar\marginnote{\bl{DG's comment}}\\}
\IfMathMode{
~~~\txt{#1}~
}{
~\\~~~#1~\\
{\normalmarginpar\marginnote{\bl{\ul{------}}}}
{\reversemarginpar\marginnote{\bl{\ul{------}}}\\}
}
}
\ClassWarning{My Macros}{#1}
}

\newcommand{\fn}[2][]{%
\IfMathMode{}{}%
\ifthenelse{\equal{}{#1}}%
{\footnote{#2}}%
{\footnote{\label{#1}#2}}%
}

\makeatletter

\newcommand{\fnref}[1]{%
\protected@xdef\@thefnmark{\ref{#1}}\@footnotemark}

\makeatother

\DeclareTextFontCommand{\bemph}{\bfseries}
\DeclareTextFontCommand{\ibemph}{\bfseries\em}
{} %
\newcommand{\e}{\emph}

{}  %

\newcommand{\bl}[1]{{\bf #1}} %

\newcommand{\il}[1]{{\it #1}} %

\providecommand{\ul}[1]{\underline{#1}} %

\newcommand{\tb}{\quad}
\newcommand{\tbb}{\qquad}
\newcommand{\tbbb}{\qquad\qquad}

\makeatother%

{}

\newcommand{\MyChangeMathMargins}{%
\setlength{\abovedisplayskip}{\abovedisplayshortskip + 5pt}%
\setlength{\belowdisplayskip}{\belowdisplayshortskip + 2pt}%
}

{}

\makeatother%

\newident{\VSP}{VSP}
\newident{\Sh}{Shape}
\newident{\Shi}{Shape_i}

\title{Entangled simultaneity versus classical interactivity\\
in communication complexity\,\fn
{
This work was presented at the 48th Symposium on Theory of Computing in 2016.
}}

\newcommand{\instDG}{Institute of Mathematics, Academy of Sciences, \v Zitna 25, Praha 1, Czech Republic.}
\newcommand{\thanksDG}{Partially funded by the grant P202/12/G061 of GA \v CR and by RVO:\ 67985840.
Part of this work was done while visiting the Centre for Quantum Technologies at the National University of Singapore, and was partially funded by the Singapore Ministry of Education and the NRF.}

{}
\author{Dmitry Gavinsky\thanks{\instDG\ \thanksDG}
}
{}

\begin{document}

\maketitle

\thispagestyle{empty}

\abstr{In 1999 Raz demonstrated a partial function that had an efficient \e{quantum two-way} communication protocol but no efficient \e{classical two-way} protocol and asked whether there existed a function with an efficient \e{quantum one-way} protocol, but still no efficient classical two-way protocol.
In 2010 Klartag and Regev demonstrated such a function and asked whether there existed a function with an efficient \e{quantum simultaneous-messages} protocol, but still no efficient classical two-way protocol.

In this work we answer the latter question affirmatively and present a partial function \Sh\ that can be computed by a protocol sending entangled simultaneous messages of poly-logarithmic size, and whose classical two-way complexity is lower bounded by a polynomial.
}

\setcounter{page}{1}

\sect[s_intro]{Introduction}

The setting of communication complexity is one of the strongest computational models where we already have tools to prove ``hardness'' -- that is, where we know problems that don't admit efficient solutions.\fn
{By calling a problem \e{hard} we always mean that \e{no efficient solution exists} (this is somewhat different from the usual notion of hardness in the context of computational complexity).}
In particular, we can compare the computational power of different communication complexity regimes (\e{classes}) via demonstrating that certain problem has an efficient solution in one regime, but not in the other.

There are three main types of the problems used for separating communication complexity classes:\ \e{total functions}, \e{partial functions} and \e{relations}.
They form a ``hierarchy'' in the following sense:\ separations via total functions are the strongest (``most convincing''), separations via relations are the weakest (and thus the easiest to obtain) and separations via partial functions are ``in between''.
There are known cases where a quantum communication complexity class can be separated from a classical one via a relation, while a functional separation is provably impossible (cf.~\cite{GRW08_Sim}).

The history of (exponential) separations in communication complexity that have demonstrated the advantage of quantum communication can be briefly outlined as follows.
\itstart
\item In 1999 Raz~\cite{R99_Exp} demonstrated a \e{partial function} that had an efficient \e{quantum two-way} communication protocol, but no efficient \e{classical two-way} protocol.
\item In 2001 Buhrman, Cleve, Watrous and de Wolf~\cite{BCWW01_Qua} 
showed that a \e{total function} (namely, the equality function) had an efficient \e{quantum simultaneous-messages} protocol \e{without shared randomness}, but no efficient \e{classical simultaneous-messages} protocol \e{without shared randomness}.
\item In 2004 Bar-Yossef, Jayram and Kerenidis~\cite{BJK04_Exp} demonstrated a \e{relation} that had an efficient \e{quantum one-way} communication protocol, but no efficient \e{classical one-way} protocol.
\item In 2008 in a joint work with Kempe, Kerenidis, Raz and de Wolf~\cite{GKKRW08_Exp}) the same separation was demonstrated via a \e{partial function}.
\item In 2008 a \e{relation} was demonstrated~\cite{G08_Cla} with an efficient \e{quantum one-way} protocol, but no efficient \e{classical two-way} protocol; the next year the result was strengthened~\cite{G09_Cla}, to admit an efficient \e{entangled simultaneous-messages} protocol.
\item In 2010 Klartag and Regev~\cite{KR11_Qua} demonstrated a \e{partial function} with an efficient \e{quantum one-way} protocol, but no efficient \e{classical two-way} protocol.
\item More recently a \e{partial function} was demonstrated~\cite{G19_Qua} with an efficient \e{quantum simultaneous-messages} protocol \e{without shared randomness}, but no efficient \e{classical simultaneous-messages} protocol, even \e{with shared randomness}.
\itend

In this work we demonstrate a \e{partial function} that can be solved by a \e{quantum simultaneous-messages} protocol that sends \asO{\log^2 n} entangled qubits; the classical two-way communication complexity of the same function is \asOm{\sq n}.
In particular, this answers an open question stated by Klartag and Regev.

A number of researchers believe that strong separations of this type are not possible for \e{total functions} -- that is, unless the model is ``too weak'' (like \e{simultaneous message passing without shared randomness}, used in~\cite{BCWW01_Qua}), it \e{might} be able to efficiently ``emulate'' classically any quantum protocol that computes a total function.
Proving or disproving this hypothesis is a major open problem.
However, if it is true, then the result of this work is very close to an ``as strong as possible'' demonstration of qualitative advantage of quantum communication over the classical one.

We note that our communication problem also has a \e{quantum one-way} protocol of cost \asO{\log^2 n} (as follows from the existence of a simultaneous-messages protocol of cost \asO{\log^2 n} that uses \asO{\log^2 n} bits of entanglement).
Till now, the only known example of super-polynomial advantage of quantum one-way over classical two-way communication in solving a functional problem has been the one demonstrated by Klartag and Regev.
Their communication problem is called \e{Vector in Subspace Problem (\VSP)} (originally proposed by Kremer~\cite{K95_Qua_Com}), and it is \e{complete} for functions\fn
{both total and partial, though not for relations!}
in quantum one-way:\ that is, the input to any function that admits an efficient quantum one-way protocol can be mapped \e{locally} to an instance of \VSP\ that \e{has the same answer as the original problem} and \e{solving which is efficient in terms of the original input size}.
Accordingly, the present work gives an alternative proof of \e{qualitative} hardness of \VSP\ for classical two-way communication\fn
{Quantitatively, the lower bound given in~\cite{KR11_Qua} for the classical two-way complexity of \VSP\ is much stronger:\ it is \asOm{n^{\dr13}}, while the following argument gives only $2^{\asOm{\sq{\log n}}}$.}%
:~if there were an efficient protocol for \VSP, there would be one for our problem as well, and we prove the opposite.

A definition of our communication problem and an overview of our approach are given in \sref[s_com]{ss_sh}, respectively.

\sect[s_prelim]{Preliminaries}

For $x\in\01^n$ and $1\le i\le n$, we will use both $x_i$ and $x(i)$ to address the \ord[i] bit of $x$.
Similarly, for $S\sbseq\set{1\dc n}$, both $x_S$ and $x(S)$ will denote the $\sz S$-bit string, consisting of naturally-ordered bits of $x$ whose indices are in $S$.
For $y\in\01^n$, let $x_y\deq x_{\set[y_i=1]i}$.
Let $\sz x$ denote the Hamming weight of $x$.
Let $x\xor y$, $x\wedge y$, $x\vee y$ and $\neg x$ denote, respectively, the bit-wise XOR, AND, OR and NOT operations.

For any discrete set $A$, let $\U[A]$ denote the uniform distribution on $A$.
Sometimes (e.g., in subscripts) we will write ``$\unin A$'' instead of ``$\sim\U[A]$''.
We will sometimes emphasise that a distribution on $\01^{2n}$ is ``viewed as bipartite'' (i.e., assumed to be the joint distribution of two random variables, containing $n$ bits each) by calling it \e{a distribution on $\01^{n+n}$}; similarly, we will write ``$(X,Y)\in\01^{n+n}$'', etc.

For a random variable $X\sim\mu$, we will use both $\h X$ and $\h\mu$ to denote the corresponding binary entropy, and similarly for the (binary) min-entropy $\hm X=\hm\mu\deq\Minn{x_0}{\log\l(\dr1{\mu(x_0)}\r)}$ and their conditional versions.\fn
{In the case of min-entropy we only allow \e{conditioning on events}, defined as the min-entropy of the corresponding distribution.}

We will often use the following property of the entropy operator:\ for any random variables $X$, $Y$ and $Z$ and event $\fb e$,
\m[m_h_non]{
\hh X{Y,\fb e} \ge \hh X{Y,Z,\fb e}
.}
We will refer to this as \e{the non-growth of entropy under conditioning}.\fn
{
Note that the entropy operator is monotonically non-increasing with respect to added conditioning \e{on variables}, but in general not with respect to conditioning \e{on events}.
}

For $p\in[0,1]$ we will denote by $\hbin p$ the \e{binary entropy function}, defined as
\m{
\hbin p \deq p\tm \log \fr1p + (1-p)\tm \log \fr1{1-p}
}
on $(0,1)$, and $\hbin0,\, \hbin1\deq0$.
Its Taylor expansion around $p=1/2$ gives
\m[m_hbin_Tay]{
1 - \hbin p
= \fr1{2\ln2}\tm\sum_{i=1}^\infty\fr{(1-2p)^{2i}}{i\tm (2i-1)}
.}

The following is a \e{weak chain rule for min-entropy}, equipped with a tail bound:

\clm[c_minch]{Let $\nu$ be a bipartite distribution on the product set $A\times B$.
Then
\mal{\E[(Y_1,Y_2)\sim\nu]{\hmm[(X_1,X_2)\sim\nu]{X_2}{X_1=Y_1}}
&\ge\hm\nu-\h[(X_1,X_2)\sim\nu]{X_1}\\
&\ge\hm\nu-\log\sz A}
and for any $\Delta\ge0$:
\m{\PR[(Y_1,Y_2)\sim\nu]
{\hmm[(X_1,X_2)\sim\nu]{X_2}{X_1=Y_1}
\le\hm\nu-\log\sz A-\Delta}
\le2^{-\Delta}.}
}

\prfstart
For all $x_1\in A$ and $x_2\in B$,
\mal{\hm\nu
&\le\log\fr1{\PR[\nu]{X_1=x_1,X_2=x_2}}
=\log\fr1{\PR[\nu]{X_1=x_1}\tm\PRr[\nu]{X_2=x_2}{X_1=x_1}}\\
&=\log\fr1{\PR[\nu]{X_1=x_1}}+\log\fr1{\PRr[\nu]{X_2=x_2}{X_1=x_1}}.}
Taking $x_2$ that minimises the second term,
\m[m_hm]{\hm\nu
\le\log\fr1{\PR[\nu]{X_1=x_1}}+\hmm[\nu]{X_2}{X_1=x_1}.}
Averaging it over all \pl[x_1] gives the first desired inequality:
\mal{\hm\nu
&\le\E[(Y_1,Y_2)\sim\nu]
{\log\fr1{\PR[(X_1,X_2)\sim\nu]{X_1=Y_1}}
+\hmm[(X_1,X_2)\sim\nu]{X_2}{X_1=Y_1}}\\
&=\h[(X_1,X_2)\sim\nu]{X_1}
+\E[(Y_1,Y_2)\sim\nu]{\hmm[(X_1,X_2)\sim\nu]{X_2}{X_1=Y_1}}.}

Averaging \bref{m_hm} over
\m{A_\Delta\deq\sett{x_1\in A}
{\hmm[\nu]{X_2}{X_1=x_1}\le\hm\nu-\log\sz A-\Delta}}
gives the tail bound:
\mal{\hm\nu
&\le\sum_{x_1\in A_\Delta}
\fr{\PR[\nu]{X_1=x_1}}{\PR[\nu]{X_1\in A_\Delta}}
\tm\l(\log\fr1{\PR[\nu]{X_1=x_1}}+\hmm[\nu]{X_2}{X_1=x_1}\r)\\
&=\hh[(X_1,X_2)\sim\nu]{X_1}{X_1\in A_\Delta}
+\log\fr1{\PR[\nu]{X_1\in A_\Delta}}\\
&\tbbb+\sum_{x_1\in A_\Delta}
\fr{\PR[\nu]{X_1=x_1}}{\PR[\nu]{X_1\in A_\Delta}}
\tm\hmm[\nu]{X_2}{X_1=x_1}\\
&\le\log\sz A+\log\fr1{\PR[\nu]{X_1\in A_\Delta}}
+\hm\nu-\log\sz A-\Delta,}
and so,
\f{\log\fr1{\PR[\nu]{X_1\in A_\Delta}}\ge\Delta,}
as required.
\prfend

We will use the following bound on the $l_1$-distance between two distributions:

\clm[c_l1-en]{Let $\nu_1$ and $\nu_2$ be distributions on $\01^n$.
Then
\m{\norm{\nu_1-\nu_2}_1^2
\le8\ln2\tm\lf(n-\Min{\h{\nu_1},\h{\nu_2}}\rt).}
}

\prf{Let $u$ be the uniform distribution on $\01^n$, then
\m{\h\mu=\sum_x\mu(x)\log\fr1{\mu(x)}
=-\sum_x\mu(x)\log\fr{\mu(x)}{2^{-n}}+\sum_x\mu(x)\log\fr1{2^{-n}}
=n-\KL\mu u.}
From the triangle and Pinsker's inequalities,
\mal{\norm{\nu_1-\nu_2}_1^2
&\le4\Maxx{j\in\12}{\norm{\nu_j-u}_1^2}\\
&\le8\ln2\tm\Maxx{j\in\12}{\KL{\nu_j}u}
=8\ln2\tm\lf(n-\Min{\h{\nu_1},\h{\nu_2}}\rt).}
}

Let $S_n$ denote the group of permutations of the set $\set{1\dc n}$, and let $\sigma_i\in S_n$ be the \ord[i] cyclic permutation (i.e., $\sigma_i(j)=i+j$ if $i+j\le n$ and $i+j-n$ otherwise).
For $x\in\01^n$ and $\tau\in S_n$, denote by $\tau(x)$ the element of $\01^n$ whose \ord[\tau(i)] position contains $x_i$ for each $i$ -- in particular, $\sigma_j(x)$ is the \f j-bit cyclic shift of $x$.

\sect[s_com]{Communication complexity}

Please see~\cite{KN97_Comm} for an extensive overview of classical communication complexity.
The quantum counterparts differ from the classical communication models in two aspects:\ the players are allowed to send quantum messages (accordingly, the complexity is measured in \e{qubits}) and to perform arbitrary quantum operations locally.
We say that a communication model allows \e{prior entanglement} if the players can share any (input-independent) quantum state and use it in the protocol (in the case of simultaneous message passing, entanglement is only allowed between Alice and Bob).

The communication problem that we use for our separation is the following partial function.

\ndefi{\Sh\ -- \e{Sh}ifted \e{Ap}proximate \e Equality}{Let Alice receive $(x_1,x_2)\in\01^{n+n}$ and Bob receive $(y_1,y_2)\in\01^{n+n}$.
Then
\m{\Sh(x_1,x_2,y_1,y_2)=\thrcase
{1}{if $\exists i:\sz{\sigma_i(x_1)\xor x_2\xor\sigma_i(y_1)\xor y_2}\le\fr{2n}5$;}
{0}{if $\forall i:\fr{7n}{15}\le\sz{\sigma_i(x_1)\xor x_2\xor\sigma_i(y_1)\xor y_2}\le\fr{8n}{15}$;}
{\txt{undefined}}{otherwise.}}
}

That is, $\Sh(x_1,x_2,y_1,y_2)$ ``asks'' whether there exists a cyclic shift $\sigma_i$, such that $\sigma_i(x_1)\xor x_2$ is close to $\sigma_i(y_1)\xor y_2$.
Note that both the ``meaning of closeness'' and the promised gap in the definition of \Sh\ are \asOm n\ -- as we will see, the former is crucial for the lower bound argument, while the latter is (apparently) essential in order to admit an efficient quantum protocol.

\ssect[ss_sh]{Intuition behind the \Sh\ and its analysis}[Intuition behind the Shape and its analysis]

To design a simultaneous-messages protocol for \Sh, we will first build a protocol for the following ``sub-problem'':
\m{\Shi(x_1,x_2,y_1,y_2)\deq\thrcase
{1}{if $\sz{\sigma_i(x_1)\xor x_2\xor\sigma_i(y_1)\xor y_2}\le\fr{2n}5$;}
{0}{if $\fr{7n}{15}\le\sz{\sigma_i(x_1)\xor x_2\xor\sigma_i(y_1)\xor y_2}\le\fr{8n}{15}$;}
{\txt{undefined}}{otherwise.}
}
Our protocol for \Shi\ will be such that the quantum messages sent by the players will be independent of $i$, and only the referee will need to know $i$ in order to measure the messages and produce the answer.
Accordingly, reducing the error of solving \Shi\ to sufficiently small (inverse-polynomial) value and sequentially applying the referee's measurements corresponding to all $0\le i<n$ to the \e{same} quantum messages received from the players will produce answers to all $n$ instances of \Shi\ (with respect to the actual input), and with high probability all these answers will be correct.
This gives a protocol for \Sh.

To show hardness of \Sh\ for classical two-way communication, we start by using relatively standard entropy-based arguments to say -- towards contradiction -- that if there is a short protocol for \Sh, then there exists a large subset $A\sbseq\01^{n+n}$, such that \e{when $(X,Y)\sim\U[A]$, ``something'' is known about $\sigma_i(X)\xor Y$ for every $0\le i<n$}, that is, \asOm1 bits of entropy are ``missing'', on average, in each of the corresponding distributions.
One possible way for $A$ to have this property would be to ``fix'' certain bits of all its elements, and the number of fixed bit-positions should be sufficient in order to have them ``overlap'' in every $\sigma_i(x)\xor y$ when $(x,y)\in A$ -- that is, roughly \asOm{\sq n} positions must be fixed (according to the ``birthday paradox''), which means that $A$ can have size at most $2^{2n-\asOm{\sq n}}$, which, in turn, is sufficient for our lower bound.

However, there is another possibility for $A$ to have the same property:\ it can to fix only the bit-parity of all its elements and have size $2^{2n-1}$, thus leading to no meaningful lower bound.
This is the reason why \Sh\ has been defined ``with margins'' (the problem asks whether the bit-wise XOR of strings is \e{close to} $\bar0$ -- not necessarily equals it):\ this way we can choose the input distribution to be ``noisy'' and draw a stronger conclusion from the existence of a short protocol, namely that \e{``something'' is known about every $\sigma_i(X)\xor \tilde Y$ when $(X,Y)\sim\U[A]$}, where $\tilde Y$ is a noisy version of the random variable $Y$.
Noise is known to ``damage'' high-degree Fourier coefficients (as quantified by the famous hypercontractive inequality) -- in particular, a set $A$ that only restricts the bit-parity of its elements would fail miserably with respect to the ``noisy condition''.

We would like to use the hypercontractivity to prove an upper bound on the size of $A$; indeed, the ``noisy condition'' is very similar to what has been analysed in~\cite{GKKRW08_Exp} and led to a good bound on $\sz A$.
There is an interesting distinction between the two cases:\ in~\cite{GKKRW08_Exp} ``something'' had to be known about $\tau(X)\xor \tilde Y$ for every $\tau\in S_n$, whereas in this work we can only require that it holds for the \pl[\sigma_i].\fn
{Instead of $\sett{\sigma_i}{0\le i<n}$ we could have fixed any other family, containing up to quasi-polynomial number of permutations; it may not be larger than that, as our quantum protocol would no longer be able to solve the problem efficiently.}
The argument in~\cite{GKKRW08_Exp} relies strongly on the symmetry resulting from allowing all $n!$ permutation -- our modestly-permuted \Sh\ seems to ask for different treatment.

There are at least two naturally-looking approaches to analyse a set $A$ that satisfies the ``noisy condition'' with respect to a small family of permutations.
First, we can use the fact that any such family corresponds to a rather small (of size roughly $2^n\tm\lra{\txt{the number of permutations}}$) subset of all $2^{2n}$ Fourier coefficients in the characteristic function of $A$ which are responsible for the entire entropy loss -- we can try to investigate the Fourier structure of Boolean functions that are ``heavily supported'' on the corresponding subset.
Second, we can use entropy-inspired arguments and try to show that staying distinguishable after the action of one of polynomially-many allowed permutations followed by noise is, essentially, as hard as staying distinguishable after the action of a uniformly-random (disclosed) permutation from $S_n$.
The proof will use the second approach.\fn
{We still use the hypercontractivity to conclude the proof, in a way somewhat similar to~\cite{KKL88_The} (cf.~\clmref{c_KKL}); in the language of the above informal description, that is done \e{after} using the presence of noise in the input distribution in order to limit (via entropy-inspired arguments) the potential role of high-degree Fourier coefficients.}

\sect[s_qua]{Solving \Sh\ with simultaneous entangled messages}[Solving Shape with simultaneous entangled messages]

Here we give a protocol for solving \Sh\ in \QIIe\ -- the model of quantum simultaneous message passing with entanglement (between Alice and Bob).

\paragraph{Protocol for \Shi:}
In the beginning Alice and Bob share two states:
$\ket{A_0}=\fr1{\sq n}\tm\sum_{k=1}^n\ket k\ket k$ and $\ket{B_0}=\fr1{\sq n}\tm\sum_{j=1}^n\ket j\ket j$,
where Alice holds the first ``register'' and Bob holds the second ``register'' of each state.
Upon receiving the input, the players apply local conditional phase-negations to transform the shared states into the form
\m{\ket{A_1}=\fr1{\sq n}\tm\sum_{k=1}^n(-1)^{x_1(k)+y_1(k)}\ket k\ket k,~~
\ket{B_1}=\fr1{\sq n}\tm\sum_{j=1}^n(-1)^{x_2(j)+y_2(j)}\ket j\ket j.}
Then the players send the parts of the shared states to the referee who applies $\sigma_{-i}$ to both registers of $\ket{A_1}$, resulting in
\m{\ket{A_2}
=\fr1{\sq n}\tm
\sum_{k=1}^n(-1)^{x_1(k)+y_1(k)}\ket{\sigma_{-i}(k)}\ket{\sigma_{-i}(k)}
=\fr1{\sq n}\tm
\sum_{k=1}^n(-1)^{x_1(\sigma_i(k))+y_1(\sigma_i(k))}\ket k\ket k.}
At this point,
\m{\bket{A_2}{B_1}
=\fr1n\tm\sum_{k=1}^n(-1)^{x_1(\sigma_i(k))+x_2(k)+y_1(\sigma_i(k))+y_2(k)}
=1-\fr2n\tm\sz{\sigma_i(x_1)\xor x_2\xor\sigma_i(y_1)\xor y_2}}
-- in particular,
\m{\sz{\bket{A_2}{B_1}}~\twocase
{\le\fr1{15}}{if $\Shi(x_1,x_2,y_1,y_2)=0$,}
{\ge\fr15}{if $\Shi(x_1,x_2,y_1,y_2)=1$.}}
The referee can distinguish the two cases with confidence $\dr12+\asOm1$ by performing the swap-test -- a two-outcome measurement that ``accepts'' with probability $\fr{1+\sz{\bket{\phi_1}{\phi_2}}^2}2$ when performed over the pair of (pure) quantum states $\ket{\phi_1}$ and $\ket{\phi_2}$.

The above protocol can be repeated \asO{\log\fr1\eps} times in parallel to bring the error down to any $\eps>0$ -- let $\Cl P_{i,\eps}$ denote the resulting protocol.
The total communication cost of $\Cl P_{i,\eps}$ is \asO{\log n\tm\log\fr1\eps} and it uses \asO{\log n\tm\log\fr1\eps} bits of entanglement.

Let $(\Pi_{i,\eps},I-\Pi_{i,\eps})$ be the \f2-outcome projective measurement that the referee applies in $\Cl P_{i,\eps}$ to the messages received from the players in order to determine the answer (with outcome $\Pi_{i,\eps}$ corresponding to answering ``$\Shi(x_1,x_2,y_1,y_2)=1$''), and let this be the only step performed by the referee.\fn
{That is, the measurement $(\Pi_{i,\eps},I-\Pi_{i,\eps})$ ``contains'' all the steps taken by the referee according to $\Cl P_{i,\eps}$.
Alternatively, we can say that the referee ``uncomputes'' his previous (recursive) steps upon having performed the (only) measurement in the end of $\Cl P_{i,\eps}$.}
Note that running $\Cl P_{i,\eps}$ didn't require either Alice or Bob to know the actual value of $i$ -- only the referee had to know it in order to apply $(\Pi_{i,\eps},I-\Pi_{i,\eps})$.
This makes $\Cl P_{i,\eps}$ a perfect ``building block'' for solving the original problem.

\paragraph{Protocol for \Sh:}
Let Alice and Bob send to the referee their messages, as prescribed by $\Cl P_{1,\eps'}$, for some $\eps'$ to be fixed soon.
The referee sequentially measures the received messages with $(\Pi_{i,\eps'},I-\Pi_{i,\eps'})$ for all $0\le i<n$.
If at least one outcome $\Pi_{i,\eps'}$ has been obtained, the referee answers ``$\Sh(x_1,x_2,y_1,y_2)=1$''; otherwise, ``$\Sh(x_1,x_2,y_1,y_2)=0$''.

Call the above protocol $\Cl P$.
Assume without loss of generality that $\Sh(x_1,x_2,y_1,y_2)\in\01$ (i.e., the input fulfils the promise).
To analyse the error of $\Cl P$, note that the protocol can return the wrong answer only if at some round $i$ the outcome of the measurement $(\Pi_{i,\eps'},I-\Pi_{i,\eps'})$ was \e{wrong} -- that is, the outcome was $\Pi_{i,\eps'}$ while $\Shi(x_1,x_2,y_1,y_2)=0$, or vice versa.
Note that while the probability of the outcome of the first performed measurement being wrong is bounded above by $\eps'$ (as follows trivially from the error bound of $\Cl P_{1,\eps'}$), at the subsequent rounds the state being measured may have been ``distorted'' by the earlier measurements, which, in turn, may increase the error probability.

We analyse\fn
{Below we give a simple proof of a sub-optimal bound, sufficient for us.
A more delicate treatment of similar setting is implicit in~\cite{ANTV02_De} and explicit in \fakelemref{2} of~\cite{A04_Lim}.}
the probability that \e{the first wrong outcome during execution of $\Cl P$ has occurred at round $j$} -- denote it by $\eps_j$.
For all $0\le i\le j$, let $\Pi_i'\in\set{\Pi_{i,\eps'},I-\Pi_{i,\eps'}}$ be the \e{right} outcome for round $i$.
Let $v$ be a unit vector representing the (pure) quantum state of the original messages received by the referee from the players (before any measurement has been applied).
Denote:
\m{\Pi_i'v = v+u_i.}
From correctness of \pl[\Cl P_{i,\eps'}] it follows that $\norm[2]{u_i}\le\sq{\eps'}$.

Let $v_j$ be the input state for the \ord[j] measurement.
As we are assuming that all the previous measurements produced the right answers,
\m{v_j=\fr{\Pi_{j-1}'\cdots\Pi_1'v}{\norm[2]{\Pi_{j-1}'\dtm\Pi_1'v}}
=\fr{v+\Pi_{j-1}'\dtm\Pi_2'u_1\dpl\Pi_{j-1}'u_{j-2}}
{\norm[2]{\Pi_{j-1}'\dtm\Pi_1'v}}
=\alpha(v+v_j'),}
where $\alpha=\dr1{\norm[2]{\Pi_{j-1}'\dtm\Pi_1'v}}\ge1$ and $\norm[2]{v_j'}\le\sum_{i=1}^{j-1}\norm[2]{u_i}<n\sq{\eps'}$.
Then
\m{\eps_j
=1-\norm{\Pi_j'v_j}_2^2=1-\alpha^2\norm{\Pi_j'v+\Pi_j'v_j'}_2^2
\le1-\norm{v\tm(1-\norm[2]{u_j}-\norm[2]{v_j'})}_2^2
<4n\sq{\eps'}.}

The error probability of $\Cl P$ is at most
$\sum_{j=1}^n\eps_j<4n^2\sq{\eps'}$ -- choosing $\eps'=\fr{\eps^2}{16n^4}$ makes it less than $\eps$.
The resulting communication cost and entanglement requirements of $\Cl P$ are $\asO{\log n\tm\log\fr1{\eps'}}=\asO{\log^2 n+\log n\tm\log\fr1\eps}$.

\theo[t_up]{The complexity of \Sh\ in \QIIe\ is \asO{\log^2 n}.}

\sect[s_cla]{Solving \Sh\ with classical interaction}[Solving Shape with classical interaction]

In this and the next sections we prove a lower bound on the complexity of \Sh\ in \R\ -- the model of interactive classical communication:

\theo[t_low]{The complexity of \Sh\ in \R\ is \asOm{\sq n}.}

The above bound is nearly-tight.\fn
{Due to the ``birthday paradox'', a single message from Alice that contains the values of \asT{\sq{n\log n}} randomly-chosen bits of $(x_1,x_2)$ allows Bob to answer $\Sh(x_1,x_2,y_1,y_2)$ with polynomially-small error probability -- using shared randomness to choose the positions, this can be implemented by a protocol of cost \asO{\sq{n\log n}}.
A similar approach can be used to get an \asO{\sq{n\log n}}-bit protocol in \RIIp\ -- the model of classical simultaneous message passing with shared randomness.}

We start by introducing several distributions.

\ndefi[def_distr]{Useful distributions -- $T_\delta$, $\mu_0$, $\mu_1$ and $\mu$}{~
\itemi{
\item For $\delta\in[0,\dr12]$, let $T_\delta$ be the distribution on $\01^n$, where each position is $1$ with probability $\delta$, independently from the other positions.
For a random variable $X\sim\nu$ taking values from $\01^n$, let $T_\delta(X)$ be the variable $X\xor Z$ and $T_\delta(\nu)$ be the distribution of $T_\delta(X)$, and for $x\in\01^n$, let $T_\delta(x)$ be the variable $x\xor Z$, where $Z\sim T_\delta$ in all cases.
\item Let $\mu_0$ be the uniform distribution on $\01^{4n}$.
\item For $0\le i<n$, let $\mu_1^{(i)}$ be the uniform distribution of $(X_1,X_2,Y_1,Y_2)$ taking values from $\01^{4n}$, modulo the condition $\sigma_i(X_1)\xor X_2\xor\sigma_i(Y_1)\xor Y_2=\bar0$, and let $\tilde\mu_1^{(i)}$ be the distribution of $(T_{\dr38}(X_1),X_2,Y_1,Y_2)$ when $(X_1,X_2,Y_1,Y_2)\sim\mu_1^{(i)}$.
Let $\mu_1$ be the distribution of sampling from $\tilde\mu_1^{(i)}$ for uniformly random $i$.
\item Let $\mu$ be the distribution of sampling $(X_1,X_2,Y_1,Y_2)\sim\mu_Z$ when $Z\sim\U[\01]$.
}}

Note that
\m[m_mu01_o1]{
\forall j\in\01:\: \PR[\mu_j]{\Sh(X_1,X_2,Y_1,Y_2)=j}\in1-\aso1
,}
and therefore,
\m{
\PR[\mu]{\Sh(X_1,X_2,Y_1,Y_2)=j}\in\dr12\pm\aso1
.}
We will show that solving \Sh\ under $\mu$ with small constant error\fn
{The support of $\mu$ is $\01^{4n}$ -- in particular, sometimes \Sh\ is undefined under $\mu$; however, this happens with probability \aso{1} and a good protocol must return \e{the} right answer from $\01$ for the given input always, except with arbitrarily small constant probability.}
requires \asOm{\sq n} bits of communication.

\subsection*{Proof outline}

Lower-bounding the \R-complexity of a communication problem is often based on showing that it admits no large nearly-monochromatic rectangles with respect to certain input distribution:\ here a (combinatorial) rectangle is a subset of input pairs of the product form $A\times B$, where $A$ and $B$ are, respectively, subsets of Alice's and Bob's inputs.
We use this approach in the analysis of \Sh.
\itstart
\item In Step 1 we show that if a short protocol were solving \Sh, then there would exist a large input rectangle $A\times B$, strongly biased towards ``$\Sh(x_1,x_2,y_1,y_2)=0$'' under $\mu$.
\item In Step 2 we say that if the above were true, then either $A$ or $B$ would ``tell something'' about each possible ``noisy shifted convolution'' of its elements (i.e., \asOm1\ bits would be known about every $\sigma_i(x_1)\xor T_{\dr14}(x_2)$, $0\le i\le n-1$ when $(x_1,x_2)$ is a uniformly-random element of the set).
\item In Step 3 we say that if the above were true, then \asOm{\sq n} bits would be known about a uniformly-random element of either $A$ or $B$.
\item In Step 4 we conclude that \theoref{t_low} holds.
\itend

\subsection*{Step 1}
If there were a short protocol solving \Sh\ with small error, then there would exist a large input rectangle, strongly biased towards ``$\Sh(x_1,x_2,y_1,y_2)=0$'' under $\mu$.
Formally:

\lem[l_rec]{Assume that a protocol of cost $c$ solves \Sh\ in \R\ with error at most $\eps$ for some $\eps\in\asOm1$.
Then for $n$ large enough, there exists a rectangle $A\times B\sbseq\01^{2n}\times\01^{2n}$ of size at least $2^{4n-c-3}$, such that
\m{\mu_1(A\times B)\le4\eps\tm\mu_0(A\times B).}
}

\prfstart
If there is a randomised protocol of cost $c$ that solves \Sh\ with error at most $\eps$, then some value of protocol's random string achieves at most the same error under the input distribution $\mu$ -- by fixing that value we obtain a \e{deterministic} protocol $\Pi$ that solves \Sh\ under $\mu$ with error at most $\eps$, while partitioning the set of possible input pairs $((x_1,x_2),(y_1,y_2))$ into (at most) $2^c$ combinatorial rectangles, each labelled with the answer that $\Pi$ gives to the corresponding input pairs.
As $\mu$ is the balanced mixture of $\mu_0$ and $\mu_1$, the error of $\Pi$ with respect to both $\mu_0$ and $\mu_1$ is at most $2\eps$.

Let $(X_1,X_2,Y_1,Y_2)$ be random variables taking the corresponding input values.
Let $\fb e_1$ be the event that $(X_1,X_2,Y_1,Y_2)$ belongs to a \pss[\Pi] rectangle of size at least $2^{4n-c-3}$.
As there are at most $2^c$ rectangles and $\mu_0$ is uniform,
\m{
\PR[\mu_0]{\fb e_1}\ge \dr78
.}
Let $\fb e_2$ be the event that $(X_1,X_2,Y_1,Y_2)$ belongs to a \pss[\Pi] rectangle $A\times B$, such that
\m{
\mu_1(A\times B) > 4\eps\tm\mu_0(A\times B)
,}
and let $q\deq\PRr[\mu_0]{\txt{$\Pi$ answers ``$0$''}}{\fb e_2}$.
Then
\m[P]{
4\eps\tm q\tm \mu_0(\fb e_2)
& = 4\eps\tm \PR[\mu_0]{\fb e_2 \wedge \txt{$\Pi$ answers ``$0$''}}\\
& \le \PR[\mu_1]{\fb e_2 \wedge \txt{$\Pi$ answers ``$0$''}}
\le \PR[\mu_1]{\txt{$\Pi$ answers ``$0$''}}\\
& \le \PR[\mu_1]{\txt{$\Pi$ errs}} + \PR[\mu_1]{\Sh(X_1,X_2,Y_1,Y_2)=0}
\in 2\eps + \aso1
,}
where the rightmost upper bound is due to~\bref{m_mu01_o1}.
Similarly,
\m{
(1-q)\tm \mu_0(\fb e_2)
& = \PR[\mu_0]{\fb e_2 \wedge \txt{$\Pi$ answers ``$1$''}}
\le \PR[\mu_0]{\txt{$\Pi$ answers ``$1$''}}\\
& \le \PR[\mu_0]{\txt{$\Pi$ errs}} + \PR[\mu_0]{\Sh(X_1,X_2,Y_1,Y_2)=1}
\in 2\eps + \aso1
.}

From the above two bounds,
\m{
4\eps\tm \mu_0(\fb e_2) \in 2\eps + \aso1
\tb\Then\tb
\mu_0(\fb e_2) \in \dr12 + \aso{\dr1\eps}
,}
and therefore,
\m{
\PR[\mu_0]{\fb e_1\wedge \neg \fb e_2}
\in \dr38 - \aso{\dr1\eps}
\sbseq \asOm1
.}
The result follows.
\prfend

\subsection*{Step 2}
If input rectangle $A\times B$ is strongly biased towards ``$\Sh(X_1,X_2,Y_1,Y_2)=0$'', then in expectation with respect to uniformly-random $i$, \asOm1\ bits are ``known'' either about $\sigma_i(X_1)\xor T_{\dr14}(X_2)$ when $(X_1,X_2)\sim\U[A]$ or about $\sigma_i(Y_1)\xor T_{\dr14}(Y_2)$ when $(Y_1,Y_2)\sim\U[B]$.
Formally:

\lem[l_inf]{Let $\mu_0$ and $\mu_1$ be distributions, as defined in~\defiref{def_distr}.
If for some $\eps\in[0,1]$ the rectangle $A\times B\sbseq\01^{2n}\times\01^{2n}$ satisfies
\f{\mu_1(A\times B)\le4\eps\tm\mu_0(A\times B)},
then
\m{\Min{
\E[i]{\h[(X_1,X_2)\unin A]{\sigma_i(X_1)\xor T_{\dr14}(X_2)}},
\E[i]{\h[(Y_1,Y_2)\unin B]{\sigma_i(Y_1)\xor T_{\dr14}(Y_2)}}}
\le n-\fr1{92}+\eps,}
where $i\sim\U[\set{0\dc n-1}]$.
}

\prfstart
By definition,
\f{\mu_1=\fr1n\sum_{i=0}^{n-1}\tilde\mu_1^{(i)},}
and therefore,
\m{\fr1n\tm\sum_{i=0}^{n-1}\tilde\mu_1^{(i)}(A\times B)\le
4\eps\tm\mu_0(A\times B).}
In particular, for some $I_0\sbseq\set{0\dc n-1}$, $\sz{I_0}\ge\dr n2$:
\m{\forall i\in I_0:
\tilde\mu_1^{(i)}(A\times B)\le8\eps\tm\mu_0(A\times B).}

Let $i_0\in I_0$.
Note that the noise operator is symmetric:\ $\PR{T_\delta(a)=b}=\PR{T_\delta(b)=a}$ for every $a$ and $b$.
Furthermore, $T_\delta(a\xor b)=T_\delta(a)\xor b$ and $T_{\delta_1}\l(T_{\delta_2}(a)\r)=T_{\delta_1(1-\delta_2)+\delta_2(1-\delta_1)}(a)$.
Therefore,
\m{\tilde\mu_1^{(i_0)}(x_1,x_2,y_1,y_2)\equiv\fr{\PR{\sigma_{i_0}(x_1)\xor T_{\dr14}(x_2)=\sigma_{i_0}(y_1)\xor T_{\dr14}(y_2)}}{2^{3n}}.}
Accordingly,
\mal{\fr{8\eps\tm\sz{A\times B}}{2^{4n}}
&=8\eps\tm\mu_0(A\times B)\\
&\ge\tilde\mu_1^{(i_0)}(A\times B)
=\sum_{\mac{(x_1,x_2)\in A\\(y_1,y_2)\in B}}
\fr{\PR{\sigma_{i_0}(x_1)\xor T_{\dr14}(x_2)
=\sigma_{i_0}(y_1)\xor T_{\dr14}(y_2)}}{2^{3n}}}
and so,
\m{\E[{\mac{(X_1,X_2)\sim\U[A]\\(Y_1,Y_2)\sim\U[B]}}]
{\PR{\sigma_{i_0}(X_1)\xor T_{\dr14}(X_2)=\sigma_{i_0}(Y_1)\xor T_{\dr14}(Y_2)}}
\le\fr{8\eps}{2^n}.}
Let $Z_1$ be $\sigma_{i_0}(X_1)\xor T_{\dr14}(X_2)$ when $(X_1,X_2)\sim\U[A]$ and $Z_2$ be $\sigma_{i_0}(Y_1)\xor T_{\dr14}(Y_2)$ when $(Y_1,Y_2)\sim\U[B]$.
Denote by $\nu_1$ and $\nu_2$ the distributions of $Z_1$ and $Z_2$, respectively.
Then
\m{\fr{8\eps}{2^n}
\ge\PR[\mac{Z_1\sim\nu_1\\Z_2\sim\nu_2}]{Z_1=Z_2}
=\E[Z\sim\nu_1]{\nu_2(Z)}.}
Therefore, $\PR[Z\sim\nu_1]{\nu_2(Z)\le\dr{16\eps}{2^n}}\ge\dr12$; let $\Cl Z_0\deq\set[\nu_2(z)\le\dr{16\eps}{2^n}]{z\in\01^n}$. 
Then
\m{\norm[1]{\nu_1-\nu_2}
\ge\sum_{z\in\Cl Z_0}\sz{\nu_1(z)-\nu_2(z)}
\ge\nu_1(\Cl Z_0)-\nu_2(\Cl Z_0)\ge\fr12-16\eps.}
By \clmref{c_l1-en},
\m{n-\Min{\h{\nu_1},\h{\nu_2}}
\ge\fr{\lf(\fr12-16\eps\rt)^2}{8\ln2}
>\fr1{23}-3\eps.}

Finally, since $\h{\sigma_i(X_1)\xor T_{\dr14}(X_2)},\h{\sigma_i(Y_1)\xor T_{\dr14}(Y_2)}\le n$ always,
\mal{&\sum_{i=0}^{n-1}\lf(n-\Min{
\h[{\U[A]}]{\sigma_i(X_1)\xor T_{\dr14}(X_2)},
\h[{\U[B]}]{\sigma_i(Y_1)\xor T_{\dr14}(Y_2)}}\rt)\\
&\tb\ge\sum_{i\in I_0}\lf(n-\Min{
\h[{\U[A]}]{\sigma_i(X_1)\xor T_{\dr14}(X_2)},
\h[{\U[B]}]{\sigma_i(Y_1)\xor T_{\dr14}(Y_2)}}\rt)\\
&\tb>\fr n2\tm\lf(\fr1{23}-3\eps\rt)}
and so,
\m{\Min{
\E[i]{\h[{\U[A]}]{\sigma_i(X_1)\xor T_{\dr14}(X_2)}},
\E[i]{\h[{\U[B]}]{\sigma_i(Y_1)\xor T_{\dr14}(Y_2)}}}
<n-\fr1{92}+\eps,}
as required.
\prfend[\lemref{l_inf}]

\subsection*{Step 3}
If \asOm1\ bits are ``known'' on average about each of $\sigma_i(X_1)\xor T_{\dr14}(X_2)$ when $(X_1,X_2)\sim\nu$, then \asOm{\sq n} bits are ``known'' about $(X_1,X_2)$ itself.
Formally:

\lem[l_bound]{If $\nu$ is a distribution on $\01^{n+n}$ for large enough $n$, such that
\m{\E[i\unin\set{0\dc n-1}]{\h[(X_1,X_2)\sim\nu]{\sigma_i(X_1)\xor T_{\dr14}(X_2)}}\le n-\delta}
for some $\delta>0$, then
\m{\hm\nu\le2n-\fr{\sq{\delta n}}{29}.}
}

A proof of the lemma is given in \sref{s_lem}.
Note that the noise operator $T$ is crucial for the statement to hold; without it, $\nu$ being the uniform distribution over all strings of even parity would provide a counterexample (this has been discussed in more detail in \sref{ss_sh}).
Note also that the bound in the conclusion is asymptotically tight, as ``entropy deficit'' of \asO{\sq n} bits is sufficient for $\nu$ to satisfy the condition (Sect.~\ref{ss_sh}).
On the other hand, the proof will not rely on any specific properties of the \pl[\sigma_i], and the condition can be somewhat relaxed by assuming sufficient entropy loss on the edge-wise XORs when averaging over a sufficiently large family of pairwise-disjoint perfect matchings between the bits of $X_1$ and $T_{\asOm1}(X_2)$.

\subsection*{Step 4}
If a protocol of cost $c$ solves \Sh\ in \R\ with error at most $\eps\le\dr1{93}$, then \lemref{l_rec} guarantees that the conditions of \lemref{l_inf} are satisfied by some rectangle $A\times B$ of size at least $2^{4n-c-3}$.
\lemref{l_inf} guarantees that either $\nu=\U[A]$ or $\nu=\U[B]$ satisfies the condition of \lemref{l_bound} for
$\delta=\fr1{92}-\eps$, which, in turn, implies that
$\Min{\sz A,\sz B}\le2^{2n-\asOm{\sq n}}$.
\theoref{t_low} follows.

\sect[s_lem]{Chasing the lost entropy}

To prove \lemref{l_bound}, we will use some tools related to Boolean Fourier transform and hypercontractivity.\fn
{Note that we are changing the meaning of the notation $\norm[p]v$:\ in \sref[s_qua]{s_cla} it stood for $\l(\sum_x\sz{v(x)}^p\r)^{\dr1p}$; now we define it to be $\l(\E[x]{\sz{v(x)}^p}\r)^{\dr1p}$, which is more common in the context of Fourier transform.}

Note that even though the noise operator $T$ is a crucial part of the lemma condition (as discussed earlier), the proof will ``handle'' $T$ with information-theoretic methods (cf.~\lemref{l_NDist}), while hypercontractivity (a technique, closely associated with the noise operator) will be used as an analytic tool for analysing the Fourier spectrum of certain ``noiseless'' probability distribution (cf.~\lemref{l_hyp}).

For $f,g:\01^n\to\RR$ and $p>0$, let $\norm[p]f\deq\lf(\E[x\unin\01^n]{\sz{f(x)}^p}\rt)^{1/p}$, $\norm[\infty]f\deq\Maxx x{\sz{f(x)}}$ and $\ip fg\deq\E[x]{f(x)\tm g(x)}$, so that $\norm[2]f^2=\ip ff$.
For  $s\sbseq\set{1\dc n}$ and $x\in\01^n$, let $\chi_s(x)\deq(-1)^{\sz{x_s}}$ and $\hat f(s)\deq\ip f{\chi_s}$.
The linear mapping $f\to\hat f$ is \e{norm-preserving} in the following sense:
$\norm[2] f^2=\sum_s\hat f(s)^2$ (\e{Parseval's identity}).

\nfct[f_BB]{Hypercontractive inequality~\cite{B70_Etu,B75_Ine}}{If $1\le p\le q$, then
\m{\norm[q]{\sum_{s\sbseq\set{1\dc n}}
\lf(\fr{p-1}{q-1}\rt)^{\dr{\sz s}2}\tm\hat f(s)\tm\chi_s}
\le\norm[p]f.}
}

From it we derive the following variations on the ``KKL theme'' (cf.~\cite{KKL88_The}).

\clm[c_KKL]{Let $f:\01^n\to\RR$, $\alpha\deq\norm[\infty]{f(x)}$ and $\beta\deq\norm[1]{f(x)}$.
Then for any $\delta\in[0,1]$:
\m{\sum_{s\sbseq\set{1\dc n}}\delta^{\sz s}\tm\hat f(s)^2\le
\alpha^2\tm\lf(\fr\beta\alpha\rt)^{\fr2{1+\delta}},}
and for any $t\le2\ln\fr\alpha\beta$:
\m{\sum_{\sz s\le t}\hat f(s)^2
\le\beta^2\tm\lf(\fr{2e\tm\ln\fr\alpha\beta}t\rt)^t.}
}

\prfstart
Choosing $q=2$ and $p=1+\delta$ in the hypercontractive inequality (\fctref{f_BB}), we get the first desired bound:
\mal{\sum_s\delta^{\sz s}\tm\hat f(s)^2
&=\norm[2]{\sum_s\delta^{\dr{\sz s}2}\tm\hat f(s)\tm\chi_s}^2\\
&\le\norm[1+\delta]f^2
=\lf(\E[x]{\sz{f(x)}^{1+\delta}}\rt)^{\fr2{1+\delta}}
=\lf(\alpha^{1+\delta}
\tm\E[x]{\sz{\dr{f(x)}\alpha}^{1+\delta}}\rt)^{\fr2{1+\delta}}\\
&\le\lf(\alpha^{1+\delta}
\tm\E[x]{\sz{\dr{f(x)}\alpha}}\rt)^{\fr2{1+\delta}}
=\alpha^{\fr{2\delta}{1+\delta}}\tm\beta^{\fr2{1+\delta}},}
where the first equality follows from Parseval's identity.

Now observe that
\m{\sum_{\sz s\le t}\delta^{\sz s}\tm\hat f(s)^2
\le\sum_{s\sbseq\set{1\dc n}}\delta^{\sz s}\tm\hat f(s)^2
\le\alpha^2\tm\lf(\fr\beta\alpha\rt)^{\fr2{1+\delta}}
\le\alpha^2\tm\lf(\fr\beta\alpha\rt)^{2-2\delta}
=\beta^2\tm\lf(\fr\alpha\beta\rt)^{2\delta},}
where the second inequality is what we have just proved.
Let $\delta\deq\fr t{2\ln(\dr\alpha\beta)}$, then
\m{\sum_{\sz s\le t}\hat f(s)^2
\le\fr{\beta^2\tm\lf(\dr\alpha\beta\rt)^{2\delta}}{\delta^t}
=\beta^2\tm\fr{\lf(\dr\alpha\beta\rt)^{\dr t{\ln(\dr\alpha\beta)}}}
{\lf(\dr t{2\ln(\dr\alpha\beta)}\rt)^t}
=\beta^2\tm\lf(\fr{2e\tm\ln\fr\alpha\beta}t\rt)^t,}
as required.
\prfend[\clmref{c_KKL}]

In the rest of this section we prove \lemref{l_bound}.

\subsection*{Proof outline}
Let us say, informally, that a distribution of bits is \e{noise-resistant} if its ``noisy version'' is distinguishable from the uniform distribution.
\itstart
\item
In Step 1 we observe that a ``projection'' of a noise-resistant distribution to a randomly-selected subset of bits can (also) be distinguished from the uniform -- that is, it cannot have the full entropy.
This argument is applied to the distribution of $\sigma_i(X_1)\xor X_2$, which is noise-resistant by assumption, letting us conclude that a ``typical'' $\l(\sigma_i(X_1)\xor X_2\r)_S$ has entropy $\sz S-\asOm\delta$.
\item 
In Step 2 we use chain decomposition over $\h{\l(\sigma_i(X_1)\xor X_2\r)_S}$ in order to ``pinpoint'' the entropy loss in $\sigma_i(X_1)\xor X_2$.
That is, as a result of the first two steps we ``trade the noise-resistance off'' for localising the entropy loss in $\sigma_i(X_1)\xor X_2$.
\item 
In Step 3 we apply hypercontractivity to conclude that $\nu$ cannot have the full entropy, quod erat demonstrandum.
\itend

\subsection*{Step 1}
If a distribution is noise-resistant, then its ``projection'' to a random $\dr23$-fraction of bits is also distinguishable from the uniform -- this will let us conclude that under the assumptions of \lemref{l_bound}, entropy loss must occur in a ``typical'' $\l(\sigma_i(X_1)\xor X_2\r)_S$ for $\sz S=\ceil{\dr{2n}3}$.
Formally:

\lem[l_NDist]{If $\nu$ is a distribution on $\01^n$, such that
$\h{T_{\dr14}(\nu)}\le n-\delta$ for $\delta>0$, then
\m{\E[\sz S=\ceil{\dr{2n}3}]{\hh[X\sim\nu]{X_S}S}
\le\ceil{\fr{2n}3}-\delta+2^{\log n-\asOm n},}
where $S$ is uniformly distributed over the subsets of $\set{1\dc n}$ of size $\ceil{\dr{2n}3}$.
}

Note that the distributions of $\sigma_i(X_1)\xor T_{\dr14}(X_2)$ and of $T_{\dr14}(\sigma_i(X_1)\xor X_2)$ are the same.
Therefore, the above lemma implies that
\m{\E[\sz S=\ceil{\dr{2n}3}]{\hh{\lf(\sigma_i(X_1)\xor X_2\rt)_S}S}
\le\ceil{\fr{2n}3}-(n-\h{\sigma_i(X_1)\xor T_{\dr14}(X_2)})+2^{\log n-\asOm n}.}
From the assumption of \lemref{l_bound} it follows that
\m[m_step1]{\E[\mac{i\unin\set{0\dc n-1}\\\sz S=\ceil{\dr{2n}3}}]
{\hh[(X_1,X_2)\sim\nu]{\lf(\sigma_i(X_1)\xor X_2\rt)_S}S}
\le\ceil{\fr{2n}3}-\delta+2^{\log n-\asOm n}.}

\prfstart[\lemref{l_NDist}]
For any $x_0\in\01^n$, $Y\sim T_{\dr14}(x_0)$ can be sampled via the following procedure.\fn
{This intuitive and convenient way of looking at the noisy distribution is used in~\cite{S15_On_th} for very similar purpose -- to analyse the noise-resistance of a Boolean function (cf.~\fakeremref{1.10} there).}
Pick independently $W,Z\sim\U[\01^n]$; on those coordinates $j$ where $W_j=1$, let $Y_j=x_0(j)$; otherwise, let $Y_j=Z_j$.
The same method can be used to sample $Y\sim T_{\dr14}(\nu)$:
\m{Y=(W\wedge X)\vee(\neg W\wedge Z),}
where $X\sim\nu$ and $W,Z\sim\U[\01^n]$ independently.

By the non-growth of entropy under conditioning~\bref{m_h_non},
\m{n-\delta\ge\h Y\ge\hh YW=\hh{X_W}W+\hh{Z_{\neg W}}W
=\hh{X_W}W+\E{n-\sz W}.}
Note that
\m{\hh{X_W}W+\E{n-\sz W}
=\sum_{k=0}^n\PR{\sz W=k}\tm\lf(\E[\sz S=k]{\hh{X_S}S}+n-k\rt)}
and that $\E[\sz S=k]{\hh{X_S}S}+n-k$ is non-increasing in $k$, as
\m{\E[\sz S=k+1]{\hh{X_S}S}-\E[\sz S=k]{\hh{X_S}S}\le1.}
Accordingly,
\mal{n-\delta
&\ge\lf(\E[\sz S=\ceil{\dr{2n}3}]{\hh{X_S}S}+n
-\ceil{\fr{2n}3}\rt)\tm\lf(1-\PR{\sz W>\ceil{\fr{2n}3}}\rt)\\
&\ge\E[\sz S=\ceil{\dr{2n}3}]{\hh{X_S}S}+n-\ceil{\fr{2n}3}
-n\tm2^{-\asOm n},}
where the last inequality follows from the Chernoff bound.
\prfend

\subsection*{Step 2}
Here we will use the chain decomposition and the non-growth of entropy under conditioning in order to ``pinpoint'' the entropy loss in $\sigma_i(X_1)\xor X_2$.

Let us look at $\E[\sz S=\ceil{\dr{2n}3}]{\hh{Y_S}S}$ for some $Y$, taking values from $\01^n$:
\mal[P]{\E[\sz S=\ceil{\dr{2n}3}]{\hh{Y_S}S}
&=\E[\tau\unin S_n]
{\sum_{i=1}^{\ceil{\dr{2n}3}}
\hh{Y_{\tau(i)}}{\tau,Y_{\tau(1)}\dc Y_{\tau(i-1)}}}\\
&=\sum_{i=1}^{\ceil{\dr{2n}3}}
\E[\tau\unin S_n]{\hh{Y_{\tau(i)}}{\tau,Y_{\tau(1)}\dc Y_{\tau(i-1)}}}\\
&=\sum_{i=0}^{\ceil{\dr{2n}3}-1}
\E[\mac{j\unin\set{1\dc n}\\S\sbseq\set{1\dc n}\smin\set j\\\sz S=i}]
{\hh{Y_j}{S,Y_S}}\\
&\ge \ceil{\fr{2n}3}\tm
\E[\mac{j\unin\set{1\dc n}\\S\sbseq\set{1\dc n}\smin\set j
\\\sz S=\ceil{\dr{2n}3}}]
{\hh{Y_j}{S,Y_S}},}
where the inequality follows from the non-growth of entropy under conditioning~\bref{m_h_non}.

Substituting $Y\deq\lf(\sigma_i(X_1)\xor X_2\rt)$ and combining the above with \bref{m_step1} (under $i\unin\set{0\dc n-1}$), we get that for large enough $n$,
\m{\E[\mac{i\unin\set{0\dc n-1}\\j\unin\set{1\dc n}\\
S\sbseq\set{1\dc n}\smin\set j\\\sz S=\ceil{\dr{2n}3}}]
{\hh[(X_1,X_2)\sim\nu]
{\lf(\sigma_i(X_1)\xor X_2\rt)_j}
{S,\lf(\sigma_i(X_1)\xor X_2\rt)_S}}
\le1-\fr\delta n.}
From the non-growth of entropy under conditioning~\bref{m_h_non} it further follows that
\mal{1-\fr\delta n\ge
&\E[\mac{i\unin\set{0\dc n-1}\\j\unin\set{1\dc n}\\
S\sbseq\set{1\dc n}\smin\set j\\\sz S=\ceil{\dr{2n}3}}]
{\hh[(X_1,X_2)\sim\nu]
{\lf(\sigma_i(X_1)\xor X_2\rt)_j}
{S,\lf(\sigma_i(X_1)\xor X_2\rt)_S}}\\
&\tbb\ge\E[i,j,S]
{\hh[\nu]{\lf(\sigma_i(X_1)\xor X_2\rt)_j}
{S, \lf(\sigma_i(X_1)\rt)_S, X_2(S)}}\\
\malabel{mal_prege}
&\tbb=\E[i,j,S]
{\hh[\nu]{X_1(\sigma_{-i}(j))\xor X_2(j)}
{S, \lf(\sigma_i(X_1)\rt)_S, X_2(S)}}
.}
For any $i_0$, at most $\floor{\dr n3}$ pairs of the form $(\sigma_{-i_0}(j), j)$ can be outside of $S_1\times S_2$ when $\sz{S_1}=\sz{S_2}=\ceil{\dr{5n}6}$.
Therefore, for uniformly-random $S$, $S_1$ and $S_2$ conditioning on $[S_1, S_2, X_1(S_1), X_2(S_2)]$ is at least as informative as conditioning on $[S, \lf(\sigma_i(X_1)\rt)_S, X_2(S)]$, and so, from the non-growth of entropy under conditioning~\bref{m_h_non} it follows that the expectation \bref{mal_prege} is at least
\m[P]{
&\E[\mac{i\unin\set{0\dc n-1}\\j\unin\set{1\dc n}\\
S_1\sbseq\set{1\dc n}\smin\set{\sigma_{-i}(j)}\\
S_2\sbseq\set{1\dc n}\smin\set j\\
\sz{S_1}=\sz{S_2}=\ceil{\dr{5n}6}}]
{\hh[\nu]{X_1(\sigma_{-i}(j))\xor X_2(j)}
{S_1, S_2, X_1(S_1), X_2(S_2)}}\\
&=\E[\mac{j_1,j_2\unin\set{1\dc n}\\
S_1\sbseq\set{1\dc n}\smin\set{j_1}\\
S_2\sbseq\set{1\dc n}\smin\set{j_2}\\
\sz{S_1}=\sz{S_2}=\ceil{\dr{5n}6}}]
{\hh[\nu]{X_1(j_1)\xor X_2(j_2)}
{S_1, S_2, X_1(S_1), X_2(S_2)}}\\
&=\E[\mac{S_1,S_2\sbseq\set{1\dc n}\\\sz{S_1}=\sz{S_2}=\ceil{\fr{5n}6}}]
{\E[\mac{j_1\unin\set{1\dc n}\smin S_1\\j_2\unin\set{1\dc n}\smin S_2}]
{\hh[\nu]{X_1(j_1)\xor X_2(j_2)}{S_1, S_2, X_1(S_1), X_2(S_2)}}}.}
So,
\m{\E[\mac{S_1,S_2\sbseq\set{1\dc n}\\\sz{S_1}=\sz{S_2}=\ceil{\fr{5n}6}}]
{\E[\mac{j_1\unin\set{1\dc n}\smin S_1\\j_2\unin\set{1\dc n}\smin S_2}]
{\hh[\nu]{X_1(j_1)\xor X_2(j_2)}{S_1, S_2, X_1(S_1), X_2(S_2)}}}
\le1-\fr\delta n,}
in particular, there exist subsets $T_1,T_2\sbseq\set{1\dc n}$ of size $\floor{\dr n6}$, such that
\m{\E[\mac{j_1\unin T_1\\j_2\unin T_2}]
{\hh[(X_1,X_2)\sim\nu]{X_1(j_1)\xor X_2(j_2)}{X_1(\wbr{T_1}), X_2(\wbr{T_2})}}
\le1-\fr\delta n,}
where $\wbr{T_i}$ stands for $\set{1\dc n}\smin T_i$.

Let $I_1\=T_1$, $I_2\=n+T_2$, $I\=I_1\cup I_2$ and $\wbr I\=\set{1\dc2n}\smin I$.
Then
\m[m_step2]{\fr\delta n
\le\E[\mac{j_1\unin I_1\\j_2\unin I_2}]
{1-\hh[Y\sim\nu]{Y_{j_1}\xor Y_{j_2}}{Y_{\wbr I}}}
\le2\tm\E[j_1\neq j_2\unin I]
{1-\hh[Y\sim\nu]{Y_{j_1}\xor Y_{j_2}}{Y_{\wbr I}}},}
where $\sz I\ge\dr n3-2$.

\subsection*{Step 3}
Finally, we apply hypercontractivity to argue that if \bref{m_step2} holds then $\nu$ cannot have the full entropy.
Formally:
\lem[l_hyp]{If $\rho$ is a distribution on $\01^n$ for large enough $n$, then
\m{\E[j_1\neq j_2\unin\set{1\dc n}]{1-\h[X\sim\rho]{X_{j_1}\xor X_{j_2}}}
\le\fr{45}{n^2}\tm\lf(n-\hm\rho\rt)^2.}
}

Using it, we can write (assuming large enough $n$):
\mal[P]{&\E[W\sim\nu]
{\lf(\sz I-\hmm[Y\sim\nu]{Y_I}{Y_{\wbr I}=W_{\wbr I}}\rt)^2}\\
&\tbbb\ge\fr{\sz I^2}{45}\tm\E[W\sim\nu]
{\E[j_1\neq j_2\unin I]
{1-\hmm[Y\sim\rho]{Y_{j_1}\xor Y_{j_2}}{Y_{\wbr I}=W_{\wbr I}}}}\\
&\tbbb\ge\fr{n^2}{406}\tm
\E[j_1\neq j_2\unin I]
{1-\hmm[Y\sim\rho]{Y_{j_1}\xor Y_{j_2}}{Y_{\wbr I}}}\\
&\tbbb\ge\fr{\delta n}{812},}
where the first inequality is an application of \lemref{l_hyp} with respect to $Y_I$ taking values from $\01^{\sz I}$, conditioned on the given value of $Y_{\wbr I}$; the second one uses the fact that $\sz I\ge\dr n3-2$; the last one follows from \bref{m_step2} and the fact that $\hm{}$ is never greater than $\h{}$.
Using \clmref{c_minch} with $A$ being the support of $Y_{\wbr I}$, $B$ being the support of $Y_I$ and $\Delta=2\log n+2$, we get
\mal{\fr1{4n^2}
&\ge\PR[W\sim\nu]{\hmm[Y\sim\nu]{Y_I}{Y_{\wbr I}=W_{\wbr I}}
\le\hm\nu-(2n-\sz I)-2\log n-2}\\
&=\PR[W\sim\nu]{\sz I-\hmm[Y\sim\nu]{Y_I}{Y_{\wbr I}=W_{\wbr I}}
\ge2n-\hm\nu+2\log n+2},}
and so,
\m{\fr{\delta n}{812}
\le\E[W\sim\nu]
{\lf(\sz I-\hmm[Y\sim\nu]{Y_I}{Y_{\wbr I}=W_{\wbr I}}\rt)^2}
\le\l(2n-\hm\nu+2\log n+2\r)^2+1.}
\lemref{l_bound} follows.

\prfstart[\lemref{l_hyp}]
For $j_1\neq j_2\in\set{1\dc n}$, let
$\Delta_{j_1,j_2}\deq\sz{\PR[\rho]{X_{j_1}=X_{j_2}}-\dr12}$.
Then
\m[m_h2del]{\E[j_1\neq j_2]{1-\h[\rho]{X_{j_1}\xor X_{j_2}}}
\le \PR[j_1\neq j_2]{\Delta_{j_1,j_2}>\dr14}
+6\E[j_1\neq j_2]{\Delta_{j_1,j_2}^2}
<22\E[j_1\neq j_2]{\Delta_{j_1,j_2}^2},}
where the first inequality follows from the Taylor expansion of the binary entropy function~\bref{m_hbin_Tay}:
\mal{
1-\h[\rho]{X_{j_1}\xor X_{j_2}}
&=\fr1{2\ln2}\tm\sum_{i=1}^\infty\fr{(2\Delta_{j_1,j_2})^{2i}}{i(2i-1)}
\le\fr{2\Delta_{j_1,j_2}^2}{\ln2}
\tm\sum_{i=0}^\infty(2\Delta_{j_1,j_2})^{2i}\\
&=\fr{2\Delta_{j_1,j_2}^2}{\ln2\tm(1-4\Delta_{j_1,j_2}^2)}
\le\fr{2\Delta_{j_1,j_2}^2}{\ln2\tm(1-2\Delta_{j_1,j_2})}
,}
where the right-hand side is at most $6\tm \Delta_{j_1,j_2}^2$ when $\Delta_{j_1,j_2}\le \dr14$, and therefore,
\m{
1-\h[\rho]{X_{j_1}\xor X_{j_2}}
\le\twocase
{6\tm \Delta_{j_1,j_2}^2}{if $\Delta_{j_1,j_2}\le \dr14$;}
{1}{always.}
}

On the other hand,
\mal{\Delta_{j_1,j_2}
&=\fr12\tm\sz{2\PR[\rho]{X_{j_1}=X_{j_2}}-1}
=\fr12\tm\sz{\E[\rho]{\chi_{\set{j_1,j_2}}(x)}}\\
&=\fr12\tm\sz{\sum_x\rho(x)\chi_{\set{j_1,j_2}}(x)}
=2^{n-1}\tm\sz{\hat\rho(\set{j_1,j_2})},}
and from \bref{m_h2del},
\m{\E[j_1\neq j_2]{1-\h[\rho]{X_{j_1}\xor X_{j_2}}}
<22\E[j_1\neq j_2]{\Delta_{j_1,j_2}^2}
<\fr{6\tm2^{2n}}{\chs n2}\tm\sum_{\sz s=2}\hat\rho(s)^2.}
From \clmref{c_KKL} with $t=2$,
\m{\sum_{\sz s=2}\hat\rho(s)^2
\le\norm[1]\rho^2\tm\lf(e\tm\ln\fr{\norm[\infty]\rho}{\norm[1]\rho}\rt)^2
=2^{-2n}\tm\lf(\fr e{\log e}\rt)^2\tm\lf(n-\hm\rho\rt)^2,}
and the result follows.
\prfend

\section{Discussion}

We have shown that the \e{partial function} \Sh\ is easy for \QIIe\ but hard for \R, which may be viewed as improving earlier understanding of \e{when quantum communication can outperform classical communication}.
Nevertheless, prior to this work there have been a number of results that have emphasised the advantage of quantum communication:\ \cite{R99_Exp,BCWW01_Qua,BJK04_Exp,GKKRW08_Exp,G08_Cla,KR11_Qua} and many others.

On the other hand, current knowledge of \e{when classical communication can ``replace'' quantum communication} -- i.e., when the advantage of using quantum communication can be at most (quasi-) polynomial in terms of complexity -- is surprisingly limited (cf.~\cite{A04_Lim,GRW08_Sim}).
We do not want to speculate now whether or not it is possible, say, to find \e{a partial function, easy for \QII~\fn[fn_QII]
{\QII\ is the model of (``unentangled'') quantum simultaneous message passing; closely related to it is \QIIp\ -- the model of quantum simultaneous message passing with shared randomness.}
but hard for \R} -- finding that out is, obviously, an important open problem; however, our intuition strongly suggests that \e{no total function can have that property}.
Proving or refuting it would be a breakthrough.

On a different note, several questions regarding the complexity of \Sh\ in other models of communication are still open.
Answering any of the following will necessarily lead to a yet-unknown (as of now) separation of communication complexity classes.
\itemi{
\item What is the \QII-complexity of \Sh?
If it has an efficient protocol, that would strengthen qualitatively the result of this work.
If it is hard, that would give the first super-polynomial separation between \QIIe\ and \QII\ via a partial function (a relational separation is known).
\item What is the \QIIp-complexity\fnref{fn_QII} of \Sh?
The consequences would be similar.
\item What is the \RIIe-complexity\fn
{\RIIe\ is the model of classical simultaneous message passing with entanglement.}
of \Sh?
The consequences would be similar, and not even a relational separation is currently known between \QIIe\ and \RIIe.
}

\phantomsection
\addcontentsline{toc}{section}{Acknowledgements}

\section*{Acknowledgements}

I am very grateful to Oded Regev, Ronald de Wolf and anonymous reviewers for many helpful comments.

\phantomsection
\addcontentsline{toc}{section}{References}

\newcommand{\etalchar}[1]{$^{#1}$}

\end{document}